# Janus-induced atomic reconstruction amplifies twist-angle modulation of interlayer thermal transport in moiré bilayers

Bin Xu[1,2,#,*], Rulei Guo[2,#], Jie Sun[2], Hiroo Suzuki[3,4*], Takuma Yoshida[4], Ichiro Nakaya[4], Koki Sawasaki[4], Yasuhiko Hayashi[3,4], Shohei Chiashi[1], Tomoki Machida[5], Junichiro Shiomi[1,2,*]

[1]Department of Mechanical Engineering, The University of Tokyo, 7-3-1 Hongo, Bunkyo, Tokyo, 113-8656, Japan

[2]Institute of Engineering Innovation, The University of Tokyo, 2-11 Yayoi, Bunkyo, Tokyo 113-8656, Japan

[3]Faculty of Environment, Life, and Natural Science and Technology, Okayama University, Okayama 700-8530, Japan

[4]Graduate School of Environment, Life, and Natural Science and Technology, Okayama University, Okayama 700-8530, Japan

[5]Institute of Industrial Science, The University of Tokyo, 4-6-1 Komaba, Meguro-ku, Tokyo 153-8505, Japan

[*]Corresponding author:
xubin@photon.t.u-tokyo.ac.jp (B. Xu)
hiroo.suzuki@okayama-u.ac.jp (H. Suzuki)
shiomi@photon.t.u-tokyo.ac.jp (J. Shiomi)
[#]These authors contribute equally

**Abstract**

In two-dimensional moiré bilayers, atomic reconstruction, the spontaneous structural relaxation toward energy-minimizing stacking registries in the near-commensurate regime, can strongly modify local stacking and interlayer coupling, providing a possibility to significantly control phonon-mediated properties. Here we show that the twist-angle dependence of interlayer thermal conductance can be modified by introducing Janus-induced mirror-symmetry breaking into bilayer $MoS_2$. The intrinsic out-of-plane dipole in $MoSSe/MoS_2$ bilayers leads to frictionless interface, and reduces the lattice deformation energy, thereby promoting atomic-reconstruction into locally distorted aperiodic moiré patterns. These features weaken interlayer coupling and suppress phonon transmission across the interface, leading to an anomalously strong twist-angle dependence of thermal conductance, with a pronounced minimum at small twist angles and a reduction rate nearly one order of magnitude larger than that of twisted bilayer $MoS_2$. Our results demonstrate that interlayer thermal transport is modified by atomic reconstruction, highlighting Janus-induced mirror-symmetry breaking as an effective way to promoting phonon engineering in two-dimensional moiré structures.

**Introduction**

The properties of crystalline materials are fundamentally governed by lattice symmetry and periodicity. In two-dimensional (2D) van der Waals (vdW) materials, these structural factors can be flexibly reconfigured through stacking and twisting, giving rise to highly tunable electronic, optical, and magnetic properties[1–3]. Particularly important is the near-commensurate regime at small twist angles, where atomic reconstruction, the spontaneous structural relaxation toward energy-minimizing stacking registries, transforms the rigid moiré geometry into a new periodic pattern composed of commensurate domains separated by disordered regions[4]. As a result, subtle differences in the initial structure can be amplified into pronounced changes in the reconstructed moiré texture and the resulting physical properties[5,6], as demonstrated in correlated electronic states[7], interlayer excitons[8,9], and spin-related phenomena[10]. Phonons, as the fundamental lattice excitations that both reflect and influence these emergent properties, have also attracted increasing attention. However, most studies have focused on static characteristics, such as mode frequencies[11] and their spatial distributions[12]. By contrast, the transport properties of phonons in 2D moiré structures remain much less explored. With recent progress in the large-area synthesis and integration of 2D devices[13,14], understanding and controlling phonon-mediated thermal transport is not only important in the fundamental phonon physics but has become increasingly important in light of their practical applications[15].

In the out-of-plane direction, weak vdW coupling strongly limits phonon transmission, in sharp contrast to the high in-plane thermal conductivity of most 2D materials. A central issue here is to elucidate how symmetry and the periodicity govern phonon transmission across atomically thin interfaces. Kim et al. demonstrated a dramatic suppression of interlayer thermal transport in randomly twisted multilayer transition metal dichalcogenides (TMD)[16], highlighting twist-angle control of moiré structures as a promising platform for phonon engineering and inspiring subsequent studies of twist-modulated interlayer thermal conductance (TC) in systems such as bilayer graphene[17], $MoS_2$ and $WS_2$[18]. A further conceptual advance came from twisted graphite interfaces, where a frictionless interfacial state was shown to suppress transverse-acoustic phonon transmission and hinder thermal transport. This work provided important insight into the correlation between the in-plane displacement energy and interlayer coupling for explaining the twist-dependent changes in TC[19].

Nevertheless, previous studies have generally shown only a gradual twist-angle dependence of TC when the stacking deviates from commensurate configurations (0° or 60° twist angles). One possible route to amplify this response is to combine twist with mirror-symmetry control through heterostructural integration. Although thermal transport across heterointerfaces has been studied [20–22], TC under the combined control of twist and hetero-structuring have only been experimentally examined in $MoS_2$/$WS_2$ bilayer[23]. Notably, integrating $MoS_2$ with $WS_2$ shows larger interlayer space than bilayer $MoS_2$. Moreover, conventional hetero-integration often reduce the extent of interlayer spacing and interlayer coupling variation[24]. In this context, conventional hetero-integration may limit the achievable enhancement in tunability.

Recent advances in atomic-scale fabrication of artificial 2D materials with broken mirror symmetry known as Janus 2D materials[25,26], like WSSe and MoSSe[27] give a possibility. The intrinsic broken mirror symmetry in the Janus monolayer can result in an out-of-plane dipole, which modifies the interlayer potential-energy landscape. This can lead to stronger out-of-plane interlayer interactions but lower interlayer friction and facilitate atomic reconstruction in the near-commensurate regime[28]. As a results, Janus-induced symmetry breaking modifies the twist-angle dependence of interlayer interactions, producing a rapid reduction and a pronounced minimum at small twist angles[29]. By altering both interlayer coupling and the reconstruction pathway of the moiré lattice, Janus functionalization provides an additional knob for enhancing control over interlayer phonon transport.

In this work, we investigate TC in twisted MoSSe/$MoS_2$ by combining twist-angle control with Janus functionalization. Time-domain thermoreflectance (TDTR) method show that TC in MoSSe/$MoS_2$ is far more sensitive to twist angle than in bilayer $MoS_2$, reaching a minimum at small twist angles. Raman spectroscopy, density functional theory (DFT) calculations, and transmission electron microscopy (TEM) reveal enhanced interlayer sliding and atomic reconstruction, leading to locally distorted aperiodic moiré patterns that drive the TC variation. This work advances the understanding of thermal transport in 2D moiré systems and demonstrates Janus functionalization as an effective phonon-engineering strategy.

**High-quality Janus vdW fabrication and structural characterization**

We fabricated twisted bilayer $MoS_2$ on a sapphire substrate using chemical vapor deposition synthesized monolayer $MoS_2$[30]. A plasma treatment was then performed to convert $MoS_2$ into MoSSe via $H_2$/Se plasma exposure (**Fig. 1a**) to synthesize twisted MoSSe/$MoS_2$. The structure of the synthesized Janus MoSSe was characterized using Raman and photoluminescence (PL) spectroscopy.

In the monolayer regions of the pristine $MoS_2$ sample, two characteristic Raman peaks ($A_1'$ and E') appear at ~404 and 384 cm$^{-1}$, respectively. The MoSSe/$MoS_2$ sample shows new peaks corresponding to the $A_1$ and E modes near 290 and 353 cm$^{-1}$, which can be attributed to the MoSSe layer. No $A_1'$ and $E'$ modes are detected in the monolayer regions in the MoSSe/$MoS_2$ sample, indicating the near-complete conversion (~100%) of $MoS_2$ to MoSSe **(Fig. 1b)**.

A PL peak is observed at ~1.85 eV in the monolayer region of $MoS_2$, and a clear redshift to ~1.75 eV is observed in the monolayer region, which is consistent with the nearly complete formation of the Janus MoSSe. In the bilayer regions, the PL features of both $MoS_2$ and MoSSe coexisted, in agreement with the Raman results. The PL intensity in the bilayer regions reduced significantly, and this could be attributed to interlayer interactions that enhance nonradiative recombination, interlayer charge transfer, and the formation of indirect bandgaps (**Fig. 1c**). These PL results confirm the successful $MoS_2$-to-MoSSe conversion by plasma and strong interlayer coupling in the vdW heterostructures.

Raman and PL maps of the MoSSe/$MoS_2$ sample (**Fig. 1d**) were used to evaluate spatial uniformity. The Raman intensity maps of the MoSSe $A_1$ mode (**Fig. 1e**) are spatially homogeneous, confirming that the entire top layer is uniformly converted to MoSSe. In contrast, the E′ mode appears only in the overlapping area, evidencing the presence of a bilayer MoSSe/$MoS_2$ (**Fig. 1f**). **Figure S1** shows the PL intensity maps of the same region at ~1.75 eV, which are uniformly detected in both the monolayer and bilayer regions. Collectively, these Raman and PL maps demonstrate successful, spatially uniform Janus conversion across the twisted bilayer sample.

We also characterized the high-frequency Raman modes of MoSSe/$MoS_2$ with different twist angles and compared them with those of the corresponding monolayer and twisted bilayer $MoS_2$ structures (**Fig. 1g-i**). In the Janus MoSSe/$MoS_2$ bilayer, the in-plane vibration mode of the MoSSe and $MoS_2$ layers (E and E′) exhibit distinct twist angle behavior. The layer-selective phonon renormalization described in *Method* section suggests that the twist-induced moiré potential couples with the intrinsic dipole of MoSSe, leading to asymmetric interlayer coupling[28]. These results are consistent with a previous study[29], which reported signatures of twist-angle-dependent interlayer interactions in bilayer MoSSe/$MoS_2$.

**Thermal conductance in twisted MoSSe/$MoS_2$**

We employed TDTR measurements for evaluating the TC of MoSSe/$MoS_2$ heterostructures with varying twist angles at room temperature (**Fig. 2a**). The effective twist angle ranged from 0 to 60° because of the inherent threefold rotational symmetry of TMD crystals. Herein, we focused on the overall TC between the Al transducer layer and sapphire substrate for the reasons discussed in *Method* section.

When the twist angle is close to 0° or 60°, TC is similar to that in the non-twisted MoSSe/$MoS_2$ synthesized from the as-grown bilayer $MoS_2$. A strong interlayer interaction in the non-twisted bilayer from the as-grown bilayer structure proves that artificially stacked interfaces in twisted bilayer are clean and strongly coupled. A decrease in TC is observed when the twist angle deviates from 0° and 60° (**Fig. 2b**). As shown by the normalized TC results **(Fig. 2c**), the twist angle tunability of TC in MoSSe/$MoS_2$ is different from the bilayer $MoS_2$ (**Fig. S2**). A remarkable minimum in TC was observed in MoSSe/$MoS_2$ within the small twist angle regime (0–5° and 60–55°), where TC reached a minimum before increasing with a further twist angle and eventually saturating in the intermediate twist angle regime (5–55°). The extent of the variation is larger than the error bar of the TDTR measurement, indicating the reliability of the nonmonotonic trend. The unique trend of TC in the Janus system aligns with the previous observation of interlayer interaction probed by Raman spectroscopy[29]. The steep change results in an extremely high twist-angle sensitivity of the TC in nearly commensurate MoSSe/$MoS_2$. The sensitivity is quantified by the reduction rate of TC within the twist angle of 0–5° before reaching the minimum. The obtained reduction rate for MoSSe/$MoS_2$ is approximately 2.68 $MWm^{-2}K^{-1}deg^{-1}$, far exceeding those observed in other twisted 2D systems, such as $WS_2$ and graphene, where TC decreases more slowly and gradually up to intermediate twist angles before saturating. Compared with bilayer $MoS_2$, Janus functionalization enhances the reduction rate by approximately one order of magnitude (0.253 $MWm^{-2}K^{-1}deg^{-1}$ for $MoS_2$). This result highlights the potential of Janus functionalization as an effective strategy for modulating TC, particularly in near commensurate stacking structures with small twist angles.

**Variation in moiré pattern under atomic reconstruction**

Given that the TC minimum occurs in the small-twist-angle regime, the corresponding evolution of the moiré pattern upon atomic reconstruction is responsible for the rapid decrease in TC. Therefore, we investigated variations in the stacking registry using TEM measurements. At intermediate twist angles, hexagonal moiré patterns are observed (**Figs. 3c, 3d**, **and S3**), consistent with previous TEM studies in bilayer $MoS_2$[31]. Similar hexagonal moiré patterns were also observed for MoSSe/$MoS_2$ at intermediate twist angles (**Figs. 3e**, **3f**, **and S4**). **Figure 4a** shows the moiré periodicity $L_0$ calculated by the rigid stacking model (see *Method*) for the bilayer $MoS_2$ and MoSSe/$MoS_2$ stacking. When the twist angle exceeds 5°, there is a negligible difference in $L_0$ between the two systems. The experimental measurements align well with the theoretical predictions for the bilayer $MoS_2$ and MoSSe/$MoS_2$ under twist angle greater than 5° (**Fig. 3a**).

For twist angles below 5°, the periodic length in bilayer $MoS_2$ deviates significantly, indicating the onset of atomic reconstruction. Corresponding moiré patterns (**Figs. 3g, h**) reveal a clear transition from hexagonal to diamond-shaped features, which is a well-known hallmark of reconstructed bilayer $MoS_2$[32]. In contrast, the moiré pattern of MoSSe/$MoS_2$ is notably different. Instead of the reconstructed diamond-like pattern observed in bilayer $MoS_2$, small-angle MoSSe/MoS2 exhibits a spatially nonuniform moiré texture with reduced long-range periodicity (**Figs. 3i, 3j**). This implies that Janus functionalization not only breaks the out-of-plane mirror symmetry but can also break the in-plane symmetry and periodicity.

The formation of this unique aperiodic pattern can be explained within the well-established framework of atomic reconstruction as a mechanism for energy compensation[11]. To this end, we performed DFT calculations. Consistent with previous work, we also found that the broken mirror symmetry of the Janus layer and the resulting intralayer dipole (**Fig. S5**) can modify the interlayer force field and the resulting stacking energy[28,29]. **Figure 3 k, i** and **Figures S6** illustrate the schematic of the bilayer TMD structures for MoSSe/$MoS_2$ and bilayer $MoS_2$ under the 0° and 60° configurations. All possible stacking types ($R_h^M$ $H_h^h$ $R_M^X$, $R_X^X$, $H_h^M$, and $H_h^X$) are modeled to explore potential reconstruction.

We then analyzed parameters influencing atomic reconstruction after Janus functionalization. In twisted bilayer systems, reconstruction expands low-energy stacking regions ($R_h^M$, $R_M^X$, $H_h^h$, $H_h^X$) at the expense of forming high-energy regions ($R_X^X$, $H_h^M$) and lattice deformation, while the residual mismatch is absorbed by narrow soliton-like domain walls (disordered regions)[11]. In this sense, larger energy differences between stacking types, lower intralayer elastic modulus, and smaller sliding energy barriers favor atomic reconstruction when transitioning from bilayer $MoS_2$ to $MoSSe/MoS_2$.

The calculations showed that the in-plane elastic modulus of MoSSe (130.5 N/m) was lower than that of $MoS_2$ (140.8 N/m), which reduced the elastic energy penalty associated with lattice distortion during reconstruction. Furthermore, we calculated the sliding energy to confirm the energy difference between the high- ($R_X^X$, $H_h^M$) and low ($R_h^M$, $R_M^X$, $H_h^h$, $H_h^X$) energy stacking configurations and sliding energy barriers between them. For the 0° stacking, stable stacking ($R_h^M$ and $R_M^X$) showed similar energies, and the energy of unstable stacking ($R_X^X$) was higher in $MoSSe/MoS_2$, indicating a larger energy difference (**Fig. 5m**). A similar trend was observed for the 60° stacking, where the energy between unstable stacking ($H_h^M$) and stable stacking ($H_h^h$ , $H_h^X$) was larger in $MoSSe/MoS_2$ (**Fig. 5n**). Sliding energy barriers appeared when the sliding energy between the two stacking configurations was higher than that of both energies. No energy barrier appeared for 0° stacking, while an energy barrier was observed in the 60° bilayer between $H_h^h$ and $H_h^X$. $MoSSe/MoS_2$ exhibits a lower sliding energy barrier of 0.0291 eV compared to that of bilayer $MoS_2$ (0.0318 eV), facilitating easier sliding. Converting this to thermal energy, it was 337 K for $MoSSe/MoS_2$, which was close to the temperature during Janus functionalization, implying that spontaneous interlayer sliding was feasible.

Although DFT calculations reveal favorable atomic reconstruction during the transition from bilayer $MoS_2$ to $MoSSe/MoS_2$, directly observing the emergence of the locally distorted moiré pattern remains challenging. Previous reports indicate that extrinsic stimuli, such as strain, can influence the moiré pattern[5], providing a possible explanation, as Janus functionalization can also be regarded as a form of extrinsic stimulation. Because the S-to-Se replacement occurs at different times in different regions, reconstruction hysteresis may impact even after the $MoS_2$-to-MoSSe transition is fully completed. A similar inhomogeneous moiré pattern was also reported in a previous study, in which a mixture of different stacking configurations was observed in non-twisted MoSSe/MoSe [33]. Consequently, the moiré pattern loses its periodicity and becomes inhomogeneous. As the reconstructed moiré pattern is highly sensitive to the initial moiré pattern and Janus functionalization process, it varies dramatically in near-commensurate $MoSSe/MoS_2$.

**Evaluation of interlayer interaction in terms of twist angle**

A more important issue is to relate this analogous moiré pattern evolution to the variation in TC. As suggested in previous work, TC is strongly governed by interlayer interactions. We therefore performed low-frequency Raman spectroscopy to probe the variations in Raman modes associated with interlayer coupling and thereby understand the unique cross-plane thermal transport behavior observed in $MoSSe/MoS_2$. **Figure 4a** show two characteristic modes in $MoSSe/MoS_2$: the shear mode located around 23 $cm^{-1}$ and the breathing mode around 33–39.3 $cm^{-1}$, which reflect the interlayer interactions along the in-plane and out-of-plane directions, respectively. Both modes were only observed when twist angles were ~0° and 60°, which is a well-established feature of twisted bilayer TMD, as elucidated in previous studies[11,34](**Fig. 4b**). Moreover, the shear mode frequency in MoSSe/$MoS_2$ is significantly lower than bilayer $MoS_2$ (**Fig. S7**), shifting from ~27.5 $cm^{-1}$ to ~22.5 $cm^{-1}$. The lower frequency of the shear mode is consistent with the TEM observations, as the shear mode is sensitive to the long-range stacking registry[11].

Simultaneously, the breathing mode exhibited a decrease in frequency as the twist angle deviated from 0 and 60°. The trends of breathing mode frequency agrees with previous experimental studies on twisted MoSSe/$MoS_2$[29] and $MoS_2$[35] (**Fig. 4c**). The frequency of the breathing mode directly reflects the second-order force constant of the out-of-plane interaction, and its evolution with the twist angle provides a sensitive probe for the interlayer interaction that varies among differently stacked samples. The magnitude of the frequency shift in MoSSe/$MoS_2$ was larger than that in $MoS_2$, particularly within the small-twist angle regime, revealing a stronger modulation of the out-of-plane interaction **(Fig. S8)**. The decay in the frequency of the breathing mode was remarkably faster within the 0–5° twist angle before reaching the minimum TC in MoSSe/$MoS_2$, which agrees with the trend observed for TC.

By plotting TC against the breathing-mode frequency of MoSSe/$MoS_2$ (**Fig. 4d**), the two quantities were found to be strongly linked, with a Pearson's correlation coefficient of 0.77. Combined with previous works[30], it reveals the role of out-of-plane interactions in determining the TC in twisted MoSSe/$MoS_2$. Moreover, the data points were distinctively divided into two groups on the plot: those with high TC and high breathing mode frequency, and those with low TC and low breathing mode frequency. We highlighted the results of the samples with a twist angle of 0–5° and 55–60° using different markers in **Fig. 4d** to identify two groups of data. The "small twist angle (SA)-A" group shows higher TC and higher breathing-mode frequency, similar to those with non-twisted samples synthesized from the as-grown bilayer $MoS_2$. In contrast, the "SA-B" group exhibits lower TC and lower breathing-mode frequency, which reflect the minimum of TC at a small twist angle near 0 and 60°. This plot confirms that the weak interlayer interaction was responsible for the rapid drop in TC within a small twist angle. It is worth noting that although the "SA-B" and medium-twist-angle samples show similar Raman features, their moiré patterns differ substantially, highlighting the fundamentally different mechanisms governing interlayer interactions and TC at small and intermediate twist angles.

Overall, considering the atomic reconstruction that occurs in the small-twist-angle regime of a twisted bilayer, Janus functionalization and the resulting dipole in MoSSe modify the stacking energy landscape and elastic modulus, leading to a dramatic reshaping of the moiré structure. This can rapidly alter the average interlayer force constants, as reflected in the pronounced changes in the breathing and shear modes. This variation in interlayer coupling is consistent with the marked reduction in interlayer thermal conductance in the Janus bilayer at small twist angles, suggesting an underlying mechanism for the dramatic change in twist-angle-dependent TC.

## Conclusion

In Janus $MoSSe/MoS_2$, a dramatic decrease in TC within the small-twist-angle regime, followed by a slight increase in the twist angle was observed. This twist angle dependent TC is extremely sensitive compared with previously reported 2D bilayer systems. Raman spectroscopy combined with TEM observations and DFT calculations show that Janus functionalization facilitates atomic reconstruction by weakening the interlayer sliding barrier, reducing the lattice deformation energy, and increasing the energy compensation for forming stable stacking configurations. Consequently, a unique locally distorted aperiodic moiré pattern emerges in twisted Janus $MoSSe/MoS_2$, which is a result of the post-reconstruction of the moiré pattern of the bilayer $MoS_2$. This aperiodic moiré pattern weakens the interlayer interaction, corresponding to a dramatic drop in TC in the small twist angle regime, while a further increase in the twist angle brings periodicity back in the moiré pattern, restoring the interlayer coupling and slightly increasing TC. Our findings advance the understanding of thermal transport mechanisms in 2D systems by linking moiré pattern engineering driven by atomic reconstruction to interlayer interactions and TC, highlighting Janus functionalization an effective engineering strategy for phonon transport.

## Methods

*Fabrication of Janus 2D materials vdW stacking*

Twisted bilayer $MoS_2$ structures were fabricated using a sequential pick-up technique based on our previous protocol[30]. Monolayer $MoS_2$ synthesized via CVD on a $SiO_2$/Si substrate was picked using a PVC/PDMS stamp through a water-assisted delamination process performed at 70 °C. We assembled twisted bilayer $MoS_2$ on a PVC film by repeating this process. Subsequently, the bilayer $MoS_2$ was transferred onto a sapphire substrate by stamping at 170 °C. The sample was placed in a vacuum desiccator and dried at 120 °C for 12 h to eliminate residual water and improve adhesion between $MoS_2$ and the sapphire substrate. Throughout the sequential pick-up process, direct contact between the $MoS_2$ surfaces and PVC stamp was avoided for maintaining a clean interface. As reference, bilayer $MoS_2$ samples grown directly by CVD were transferred onto sapphire substrates without intentional twisting. After transfer, the samples were annealed at 320 °C for 12 h in a forming gas atmosphere ($H_2$ ~4%/Ar) to promote atomic reconstruction and enhance interlayer coupling.

Then, plasma treatment was performed to selectively substitute the uppermost sulfur atoms with Se atoms, converting bilayer $MoS_2$ into a Janus MoSSe/$MoS_2$ heterostructure. The plasma reaction was conducted in a 25-mm-diameter quartz tube equipped with an inductively coupled RF coil located upstream. The chamber was evacuated to a base pressure of 1.0 Pa, followed by the introduction of $H_2$ gas (100 sccm) to maintain a working pressure of 140 Pa. Plasma was generated at an RF power of 20 W. Bilayer $MoS_2$ samples and selenium powder (~2 g) were separately loaded into 77-mm-long quartz boats placed downstream and upstream, respectively, so that they were positioned next to each other. The growth substrate was positioned ~2 cm upstream from the plasma tip. Plasma-assisted Se substitution was performed for ~10 min. After the fabrication, the twist angle was determined by the optical microscope observation from the relative orientation of stacked triangular $MoS_2$ flakes, whose crystal axes align with their edges (**Fig. S9**).

*Raman/PL characterization*

High-frequency Raman measurements were performed using an inVia system (Renishaw) equipped with a 532 nm laser and 3000 grooves/mm grating with a wavenumber accuracy of 0.02 $cm^{-1}$. A 50× objective lens was used to focus the laser beam onto a spot of ~500 nm. The laser power was maintained at 1.91 mW to minimize laser-induced heating effects and prevent Raman peak shifts, and no observable peak shifts were detected under these conditions. The sapphire Raman peak was used to calibrate the spectra and correct any residual heating effects caused by the laser.

PL measurements were conducted using the same setup, with the grating replaced with 600 grooves/mm to enhance signal intensity. Raman and PL mapping measurements were performed on this system, with spatial step sizes set to either 1 or 2 μm depending on the resolution requirement.

Low-frequency Raman spectra were acquired using a custom-built setup equipped with a 488 nm laser and 2400 grooves/mm grating. A 50× objective lens was used for focusing and the laser power was attenuated to 3.5 mW using a neutral density (ND) filter to avoid heating effects. Two notch filters were installed in series before the spectrometer to enhance the detection of low-frequency vibrational modes. This optical configuration enabled the measurement of low-frequency Raman signals down to 10 $cm^{-1}$, with a wavenumber resolution of 0.1 $cm^{-1}$.

*TDTR measurement*

The TDTR technique was employed to measure the TC between Al transducers and sapphire substrates mediated by twisted bilayer structures. Following the fabrication and structural characterization of the Janus vdW stacking, a 100-nm-thick Al layer was deposited on top of the samples via electron beam (EB) evaporation under a high vacuum of less than $10^{-3}$ Pa. EB evaporation, which is a gentler deposition method than sputtering, is utilized with a large source-to-substrate distance to minimize the thermal effects on the sample surface during Al deposition. In addition, interlayer interactions established during annealing are largely preserved throughout the deposition process because the vdW structure samples are annealed at elevated temperatures prior to metal deposition to optimize the interlayer binding and stabilize the bilayer structures.

For the TDTR setup, an 80 MHz repetition rate laser with a pulse width of 140 fs is used. The pump and probe beams are focused to spot diameters of ~8 μm (400 nm wavelength) and 3 μm (800 nm wavelength), respectively. The pump beam is modulated at a frequency of 10 MHz, which facilitates shallow optical penetration and enhances the sensitivity to TC at the Al/sapphire interface, as confirmed by the sensitivity analysis.

The temperature decay signal ($-V_{in}/V_{out}$) is recorded over a delay time range from 0–8 ns by adjusting the mechanical delay stage. The data analysis is restricted to delay times greater than 200 ps to ensure the accurate extraction of TC values, minimizing the contributions from electron–phonon coupling. Thermal model fitting incorporated bulk values for the specific heat capacity and thermal conductivity of both the metal and dielectric layers. The thickness of the Al transducer (~100 nm) was verified through independent measurements of the reference Al/quartz samples.

In the 2D system, the overall TC ($G$) between the Al transducer layer and sapphire substrate was measured in TDTR. Sensitivity analysis (**Fig. S10**) confirmed that $G$ exhibits sufficient sensitivity to ensure the reliability of the TDTR measurements. The temperature decay profile and theoretical fitting curve for representative samples are shown in **Fig. S11**. Here, $G$ comprises three distinct components for the bilayer TMD: the thermal boundary conductance between Al and the TMD layer ($G_{\text{Al–TMD}}$), the cross-plane TC of the bilayer TMD ($G_{\text{TMD}}$), and the thermal boundary conductance between the TMD and sapphire substrate ($G_{\text{TMD–Sap}}$).

As reported in previous work, the $G_{\text{TMD}}$ of the bilayer $MoS_2$ samples can be obtained using a series resistance model by subtracting the contributions of $G_{\text{Al–TMD}}$ and $G_{\text{TMD–Sap}}$, which are independently obtained from TDTR measurement of monolayer $MoS_2$. However, for $MoSSe/MoS_2$ heterostructures, the presence of asymmetric interfaces prevents the direct extraction of $G_{TMD}$ using the same approach. However, $G_{Al-TMD}$ and $G_{TMD-Sap}$ remains constant across bilayer samples with different twist angles because of the drop-off of 2D materials onto substrates; further, Al deposition is performed under identical condition for all samples. Consequently, we focused on the trend in $G$ to discuss the twist angle dependence of the out-of-plane thermal conductance.

*TEM measurement and analysis*

TEM measurements were conducted to examine moiré patterns and atomic configurations of the bilayer $MoS_2$ and $MoSSe/MoS_2$ stacked structures. the bilayer samples were transferred onto Cu TEM grids coated with holey carbon films. This transfer ensured that the suspended regions of heterostructures were accessible for high-resolution imaging.

The measurements were performed using a field-emission transmission electron microscope (JEM-ARM200F (Dual), JEOL) operated at 80 kV. All observations were performed under low-dose conditions whenever applicable to minimize beam-induced structural modifications and damage. Under these mild observation conditions, no visible drift or lattice damage was observed during imaging.

Both bright-field and high-resolution TEM (HRTEM) modes were used to resolve the moiré superlattice and identify local atomic arrangements. An FFT analysis of the TEM images was performed to determine the relative rotational angle and confirm the crystallinity of the stacked layers. The color of the TEM image was treated using ImageJ software to improve visibility, and no filtering or masking was applied. The twist angle was measured by analyzing the angle between the two hexagonal FFT patterns. However, moiré patterns and FFT analyses alone are insufficient for unambiguously identifying bilayer TMDs with twist angles of 0° and 60°, particularly when using conventional HRTEM. Therefore, the twist angle mentioned in the TEM section refers to the twist angle difference between either 0 or 60° twist stackings.

The rigid stacking model is applied for interpreting the TEM images. This model assumes ideal 2D stacking without lattice deformation and is reliable for estimating the periodicity of twisted bilayer systems, especially at intermediate twist angles, where atomic reconstruction is negligible. For twisted bilayer 2D materials, the moiré periodicity can be estimated using $L_0 = \frac{a}{\sqrt{\theta^2+\delta^2}}$, where $a$, $\theta$, and $\delta$ represent the lattice constant of monolayer TMD, twist angle, and lattice mismatch between the upper and bottom layers, respectively[36]. The lattice constants determined from the DFT calculations are 0.312 nm for $MoS_2$ and 0.318 nm for MoSSe. This model is used to estimate the periodicity in TEM measurement, so as to confirm the possibility of atomic reconstruction.

*DFT calculation*

First-principles DFT calculations were performed using the Vienna Ab initio Simulation Package[37]. The projector augmented-wave method[38] was employed to describe interactions between the ions and electrons. The local density approximation[39–42] was adopted for the exchange-correlation potential. A plane-wave energy cut-off of 520 eV was used for all calculations.

Initial structural relaxation was performed for the untwisted (0°) configuration. The convergence criteria for electronic self-consistency and ionic relaxation were set to $10^{-8}$ eV and $10^{-6}$ eV/Å, respectively. Subsequent twisted structures were generated using the CellMatch software[43], ensuring that the strain introduced by lattice matching was less than 0.1% and the total number of atoms in the primitive cell was kept below 120. In SI, we also built the model for twist angles of 0, 13, 22, 28, 32, 38, 47, and 60° whose results is discussed in SI. For each configuration, the interlayer binding energy, total charge density, differential charge density[44], and Bader charge[45]distribution were utilized to analyze the effects of the twist angle on interlayer interaction.

We used the method described in Ref.[35] to calculate the Raman shift frequency, so as to confirm the reliability by comparing with the experimental observation. Further, the sliding energy was calculated using the nudged-elastic band method[46] to analysis the energy variation for the atomic reconstruction. The elastic modulus was calculated using the energy-strain method with the VASPKIT code[47,48].

**Acknowledgements**

This project was supported by JST-CREST (grant number JPMJCR21O2), JST-Mirai (grant number JPMJMI21G9), and JSPS KAKENHI (grant numbers JP22H04950, JP24K00817, JP25K01624, and JP24H01197). B.X. acknowledges the support from the ATI Foundation. H.S. acknowledges support from the Shorai Foundation for Science and Technology, Hirose Foundation, and Kao Foundation for Arts and Sciences.

**Author contributions**

B.X. and R.G. contributed equally to this study.

B.X. and J.S. conceived and designed the experiments; R.G. and J.S. performed the DFT simulations; B.X., T.M., K.S., H.S., and Y.H. fabricated the samples; B.X. and S.C. performed the characterization; B.X. performed the measurements with support from R.G. B.X., R.G., and J.S. wrote the manuscript; and J.S. supervised the research.

The manuscript was written with contributions from all authors. All authors approved the final version of the manuscript.

**Conflict of interest**

The authors declare no conflict of interest.

## Figure

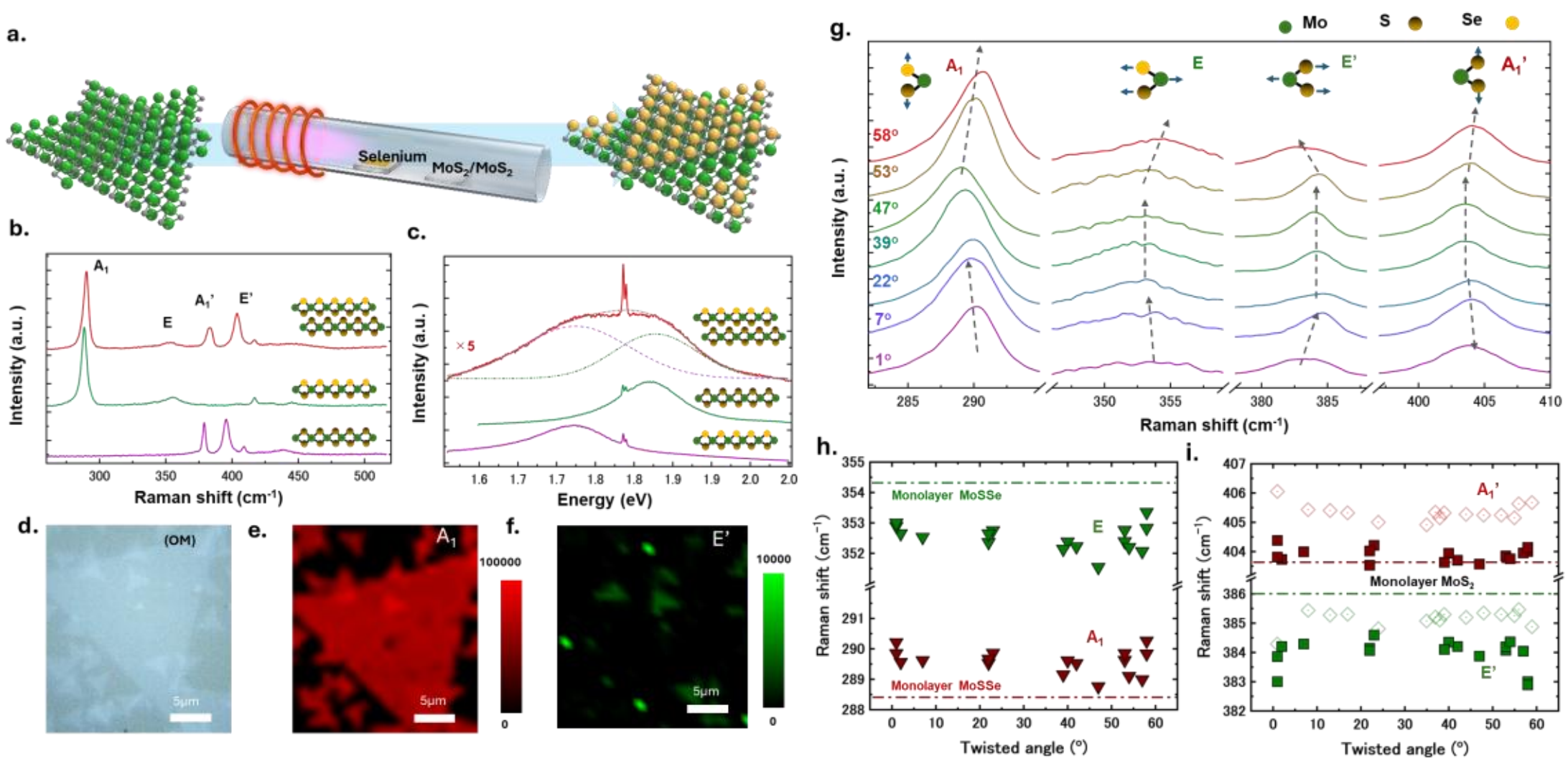

**Figure 1. Fabrication of high-quality twisted bilayer MoSSe/$MoS_2$. a.** Schematic of plasma treatment that transforms the twisted bilayer $MoS_2$ to MoSSe/$MoS_2$. **b.** Raman and **c.** PL spectra of monolayer $MoS_2$, MoSSe, and MoSSe/$MoS_2$ bilayer. The PL intensity of MoSSe/$MoS_2$ bilayer is magnified five times. **d.** Optical microscopy image and corresponding Raman mapping of **e.** $A_1$ and **f.** E' modes at identical region. **g.** Raman spectrum of MoSSe/$MoS_2$ bilayer with different twist angle at a high frequency range (280–410 $cm^{-1}$). The variation of **h.** E and $A_1$ peaks in the twisted bilayer MoSSe/$MoS_2$ and the **i.** $A_1$', E' peaks in the twisted bilayer $MoS_2$ (hollow diamonds) and MoSSe/$MoS_2$ (solid squares).

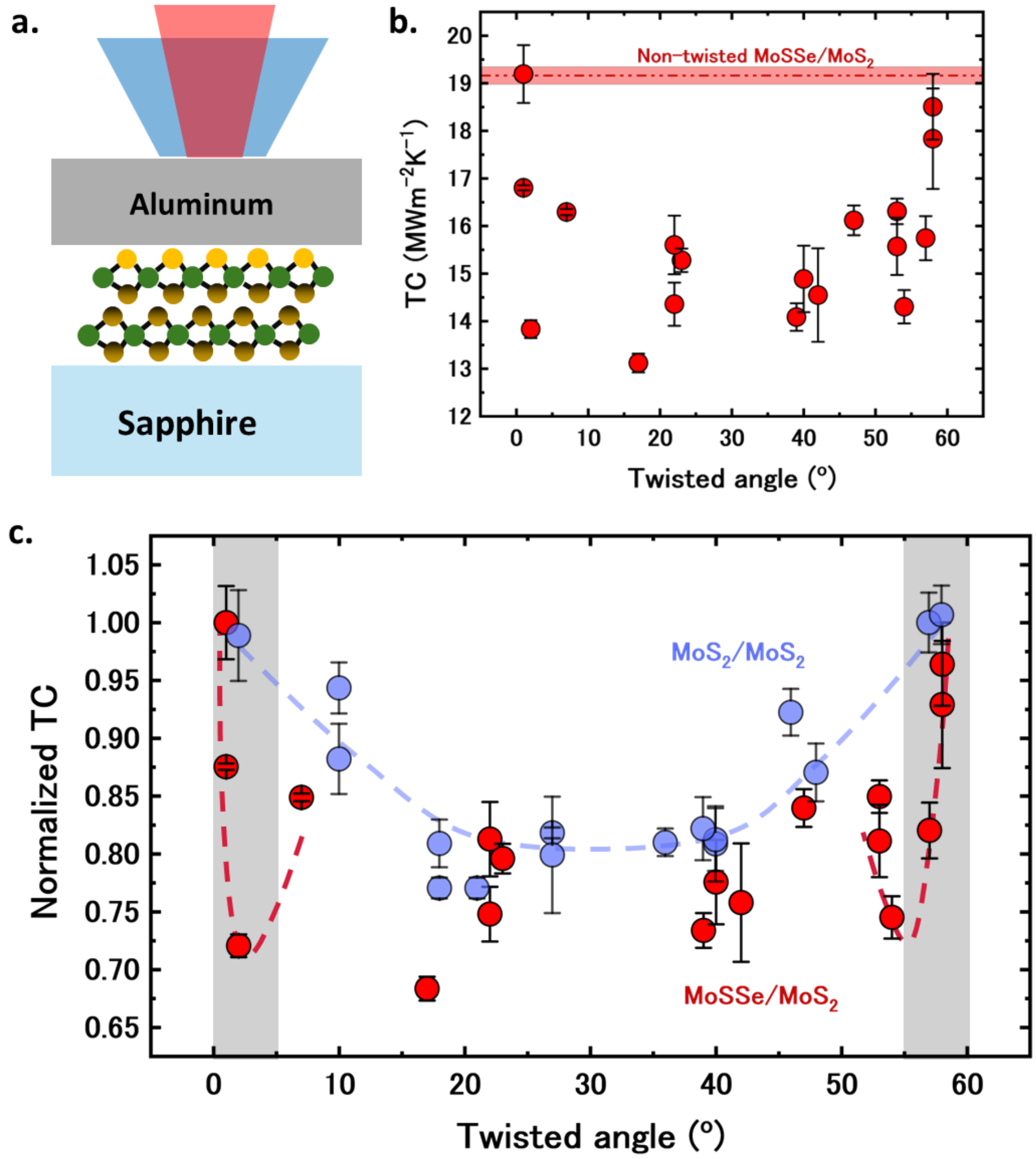

**Figure 2. Interlayer thermal conduction measurement. a.** Schematic of TDTR measurement for TC measurement. **b.** Total TC between aluminum and sapphire with varying twist angles in the $MoSSe/MoS_2$ bilayer. **c.** Normalized TC of twisted bilayer $MoS_2$ and $MoSSe/MoS_2$ under different twist angles; Red dash curves are provided as guides for viewing the pronounced minimum in TC for $MoSSe/MoS_2$ at small twist angle regime, and purple dash curves are for the slow decay in TC for bilayer $MoS_2$; grey aera represents the small twist angle regime.

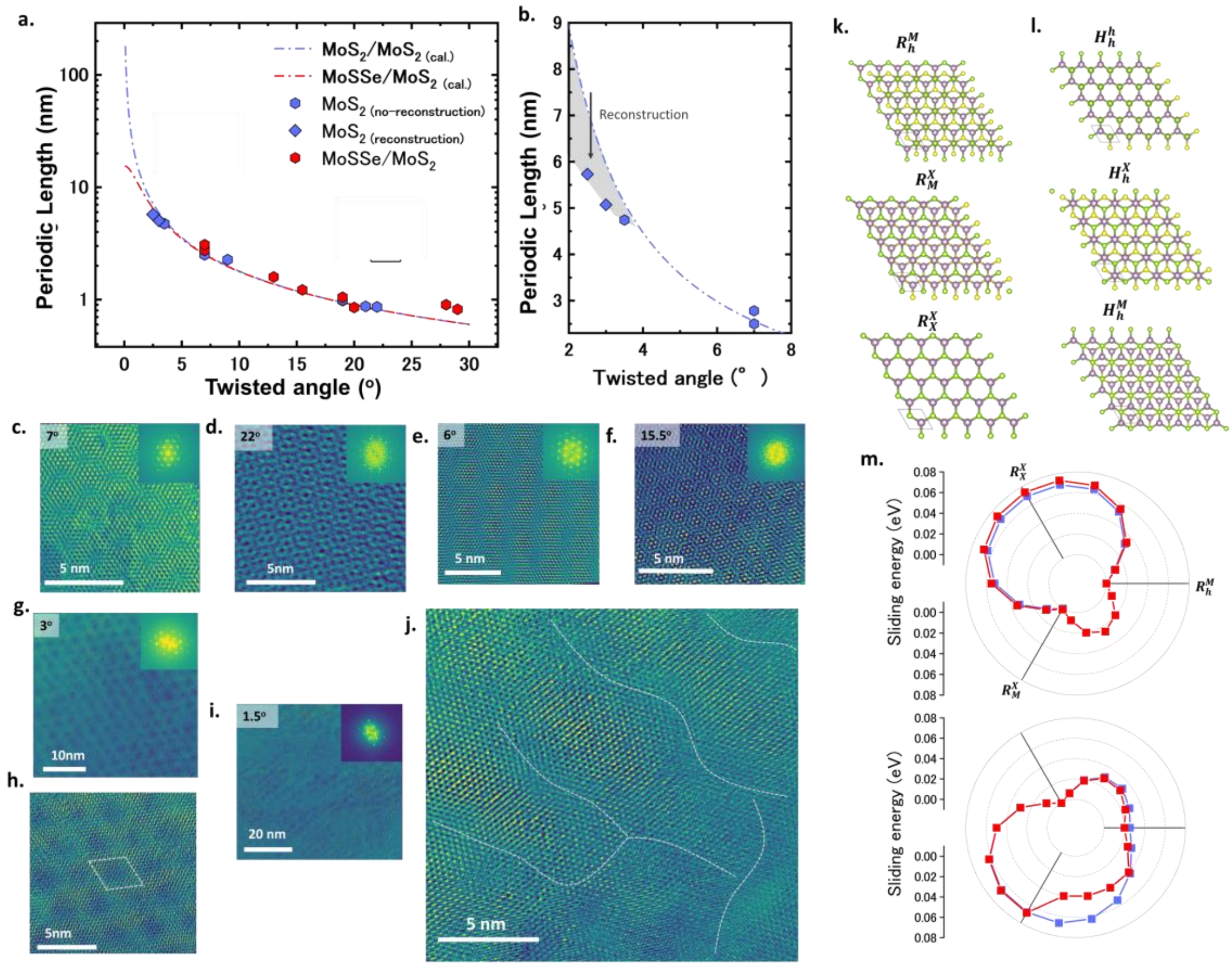

**Figure 3. Moiré pattern evolution in twisted bilayer TMD. a.b** Periodic length of the moiré pattern; curves represent the theoretical calculation based on the rigid stacking model, while mark corresponds to experimental results. TEM image of $MoS_2$ bilayer with **c, d** intermediate and **g, h** small twist angle. TEM image of MoSSe/$MoS_2$ with **e, f** intermediate and **i, j** small twist angle; **h** and **j** are magnified view for **g** and **i**, respectively. Inset are the FFT of TEM image; Dash line in **h** and **j** are the guide for eye for the disordered boundary of individual commensurate or near-commensurate domines. Representative stacking configurations for **k.** 0° and **l.** 60° individually in MoSSe/$MoS_2$. **m. n.** Sliding energy between different stacking configurations; Notably, there are only two kinds of stacking configurations for the 0° $MoS_2$ bilayer; red mark for MoSSe/$MoS_2$ and purple mark for bilayer $MoS_2$.

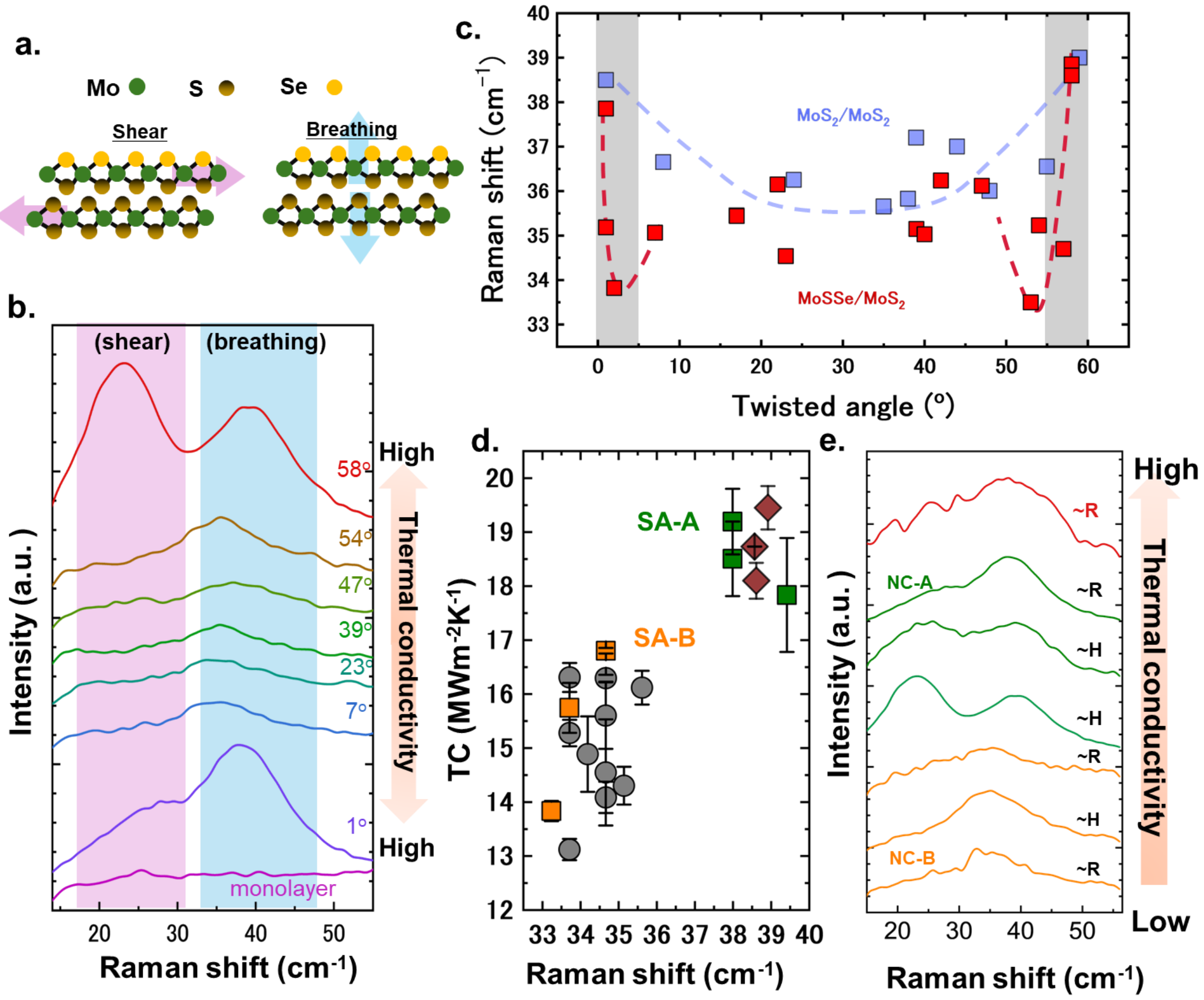

**Figure 4. Low frequency Raman modes of twisted bilayer TMD. a.** Schematic of shear mode and breathing mode. **b.** Representative low-frequency Raman spectrum of bilayer $MoSSe/MoS_2$ with varied twist angle from 1–58º. **c.** Frequencies of breathing modes in bilayer $MoS_2$ and $MoSSe/MoS_2$ under different twisted angles; Red dash curves are provided as guides for viewing the pronounced minimum in breathing mode frequency for $MoSSe/MoS_2$ at small twist angle regime, and purple dash curves are for the slow decay in breathing mode frequency for bilayer $MoS_2$; grey aera represent the small twist angle regime. **d.** Thermal conductance (TC) of $MoSSe/MoS_2$ samples with different breathing modes frequencies; grey circles for the intermediate twist angle (5–55°), orange squares for the small twist angle (0–5° and 55–60°) samples with low TC (SA-B) and green squares for the small twist angle samples with high TC (SA-A), red diamond for the R stacking samples fabricated from the as-grown bilayer $MoS_2$. **e.** Low-frequency Raman spectrum of the small twist angle SA-A and SA-B samples.

# Supplementary information: Janus-induced atomic reconstruction amplifies twist-angle modulation of interlayer thermal transport in moiré bilayers

Bin Xu[1,2,#,*], Rulei Guo[2,#], Jie Sun[2], Hiroo Suzuki[3,4*], Takuma Yoshida[4], Ichiro Nakaya[4], Koki Sawasaki[4], Yasuhiko Hayashi[3,4], Shohei Chiashi[1], Tomoki Machida[5], Junichiro Shiomi[1,2,*]

[1]Department of Mechanical Engineering, The University of Tokyo, 7-3-1 Hongo, Bunkyo, Tokyo, 113-8656, Japan

[2]Institute of Engineering Innovation, The University of Tokyo, 2-11 Yayoi, Bunkyo, Tokyo 113-8656, Japan

[3]Faculty of Environment, Life, and Natural Science and Technology, Okayama University, Okayama 700-8530, Japan

[4]Graduate School of Environment, Life, and Natural Science and Technology, Okayama University, Okayama 700-8530, Japan

[5]Institute of Industrial Science, The University of Tokyo, 4-6-1 Komaba, Meguro-ku, Tokyo 153-8505, Japan

[*]Corresponding author:

xubin@photon.t.u-tokyo.ac.jp (B. Xu)

hiroo.suzuki@okayama-u.ac.jp (H. Suzuki)

shiomi@photon.t.u-tokyo.ac.jp (J. Shiomi)

[#]These authors contribute equally

**High-frequency Raman and phonon frequency mismatch in $MoSSe/MoS_2$**

High-frequency Raman peaks correspond to optical phonons near the Γ-point, and thus serve as an indicator of the phonon energy landscape in the vicinity of the interface. As the twist angle deviates from 0° and 60° stacking, the out-of-plane $A_1$ mode shifts slightly upwards to ~288.7–290.3 $cm^{-1}$ from the monolayer MoSSe value of 288.4 $cm^{-1}$, while the in-plane E mode shifts downward to ~351.5–353.3 $cm^{-1}$ from 354.4 $cm^{-1}$. Both the $A_1$ and E modes show a slight downward trend (**Fig. 1h**). The out-of-plane $A_1'$ mode originating from the underlying $MoS_2$ layer lies in the range of 403.3–404.4 $cm^{-1}$, comparable to that of monolayer $MoS_2$, and exhibits a smaller variation (~1.1 $cm^{-1}$) than in bilayer $MoS_2$, where the $A_1'$ mode shifts upward by ~1.7 $cm^{-1}$ (405.0–406.7 $cm^{-1}$) due to twist angle effects. The in-plane $E'$ mode spans ~382.9–384.5 $cm^{-1}$, which is lower than the monolayer value and shows a larger variation (~1.6 $cm^{-1}$) than in the $MoS_2$ bilayer case (~0.8 $cm^{-1}$, from 385.5 to 384.7 $cm^{-1}$). The lower frequencies of both $A_1'$ and $E'$ in $MoSSe/MoS_2$ compared with bilayer $MoS_2$ indicate overall lattice softening of the underlying $MoS_2$ layer (**Fig. 1i**). These results reported a clear signature of twist-angle-dependent interlayer interactions in bilayer $MoSSe/MoS_2$.

Compared to bilayer $MoS_2$, the introduction of the heavier Se atoms leads to a significant reduction in both out-of-plane ($A_1$) and in-plane (E) phonon frequencies compared to $MoS_2$ ($A_1'$ and $E'$). This is consistent with the mass effect and altered bonding conditions, which shifts the phonon density of state (pDOS) of MoSSe layer towards lower frequencies. Particularly, the out-of-plane $A_1$ ($A_1'$) mode exhibits a frequency difference exceeding 100 $cm^{-1}$ between MoSSe and $MoS_2$, which is much larger than the shift observed in the in-plane E ($E'$) mode. This Raman frequency mismatch reflects the pDOS mismatch between $MoS_2$ and MoSSe, which hinders the cross-interface phonon transport. In contrast, the associated frequency changes (~a few $cm^{-1}$) with respect to twist angle are much smaller. In this sense, the twist angle should be less effective as a control parameter for pDOS matching in Janus heterostructures than in homo-bilayers.

The Pearson correlation coefficients between the TC and the high frequency Raman peaks positions of $MoSSe/MoS_2$ bilayer under different twist angles are shown in **Fig. S12**. Compared to the breathing mode, the high frequency Raman peaks are less correlated with TC, with the correlation coefficients found to be –0.68, 0.64, 0.60, and 0.55, corresponding to the $E'$, $A_1'$, $A_1$, and E modes, respectively. Besides, despite the distinctive difference in low frequency breathing mode, no clear trend was observed for the peak shift of high frequency Raman modes in SA-A and SA-B types $MoSSe/MoS_2$ (**Fig. S13**). Overall, the changes in phonon density of states (pDOS), provide less important influence in determining the interlayer phonon transport.

**Enhanced twist-angle dependence of charge distribution via Janus modification**

We investigated bilayer heterostructures composed of $MoS_2$/$MoS_2$ and MoSeS/$MoS_2$ with varying twist angles. Twist angles of 0, 13, 22, 28, 32, 38, 47, and 60° were considered. **Figures S14–S16** illustrate the schematic of the bilayer TMD structures, and the supercells adopted for MoSSe/$MoS_2$ and bilayer $MoS_2$ at these representative twist angles

On transitioning from $MoS_2$ to MoSSe, a built-in out-of-plane dipole is observed (**Fig. S5**), with the charge density more strongly localized near S atoms in the MoSSe layer. To confirm the reliability of the DFT calculation, we compared the DFT-calculated Raman shift with the experiment, which shows good agreement (**Fig. S17**). The spillover of intralayer dipoles into the adjacent $MoS_2$ layer is confirmed by the Bader net atomic charge analysis (**Fig. S18**). Compared with bilayer $MoS_{2,}$ where Mo atoms in the top and bottom layers exhibit similar Bader net charges of 1.085–1.10, MoSSe/$MoS_2$ structure shows Bader net charges of 0.945–0.965 for the top Mo atom and 1.085–1.11 for the bottom Mo atom in the $MoS_2$ layer, deviating from those in the bilayer $MoS_2$. This difference in the Bader net atomic charge of the bottom Mo layer serves as direct evidence of the effect of the intralayer dipole on the interlayer interaction. Further, we observed a stronger twist angle dependence of the Bader net charge in the MoSSe/$MoS_2$ heterostructure than that in the bilayer $MoS_2$, which indicates that the Janus dipole enhances the sensitivity of interlayer interactions to the twist angle. The modification of interlayer interactions caused by the spillover of the intralayer dipole to the interlayer significantly reduces interlayer spacing (**Fig. S19**), leading to higher sensitivity to the out-of-plane interlayer force than that observed for bilayer $MoS_2$.

**Supporting Figures**

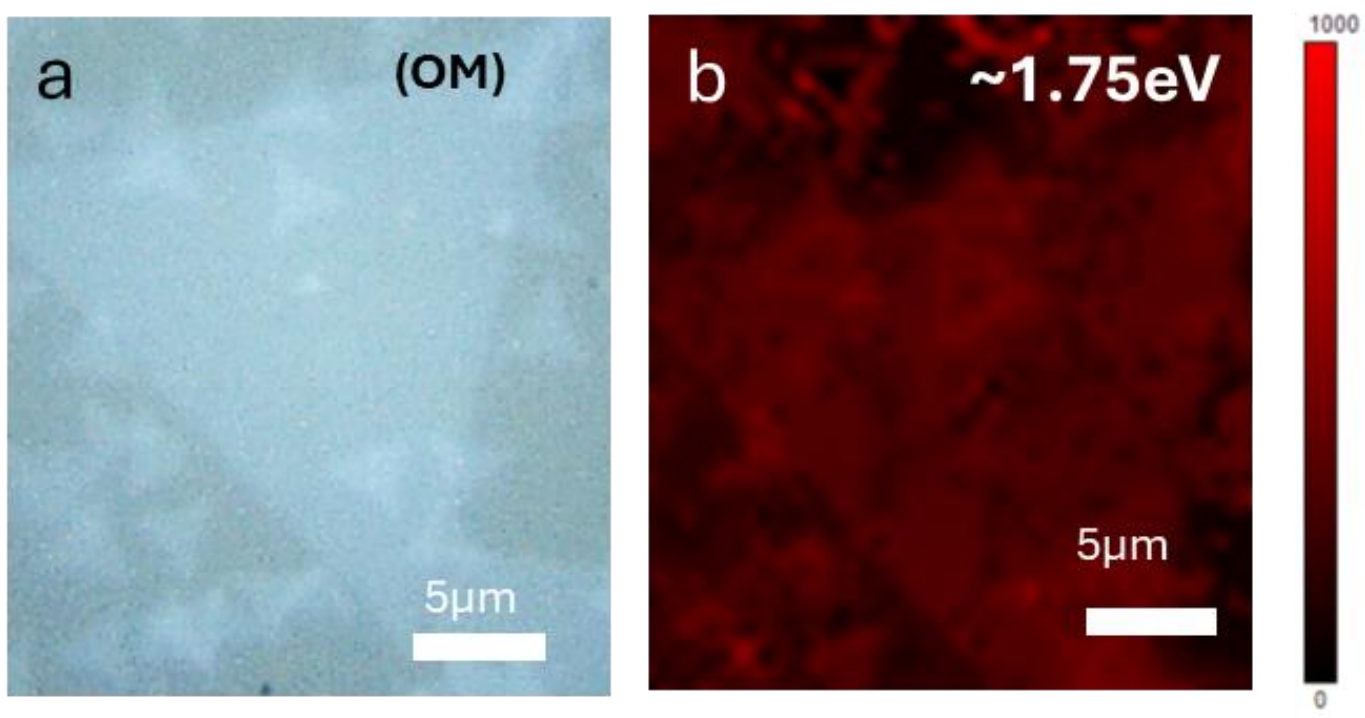

**Figure S1.** PL mapping result of MoSSe (Intensity at ~1.75eV).

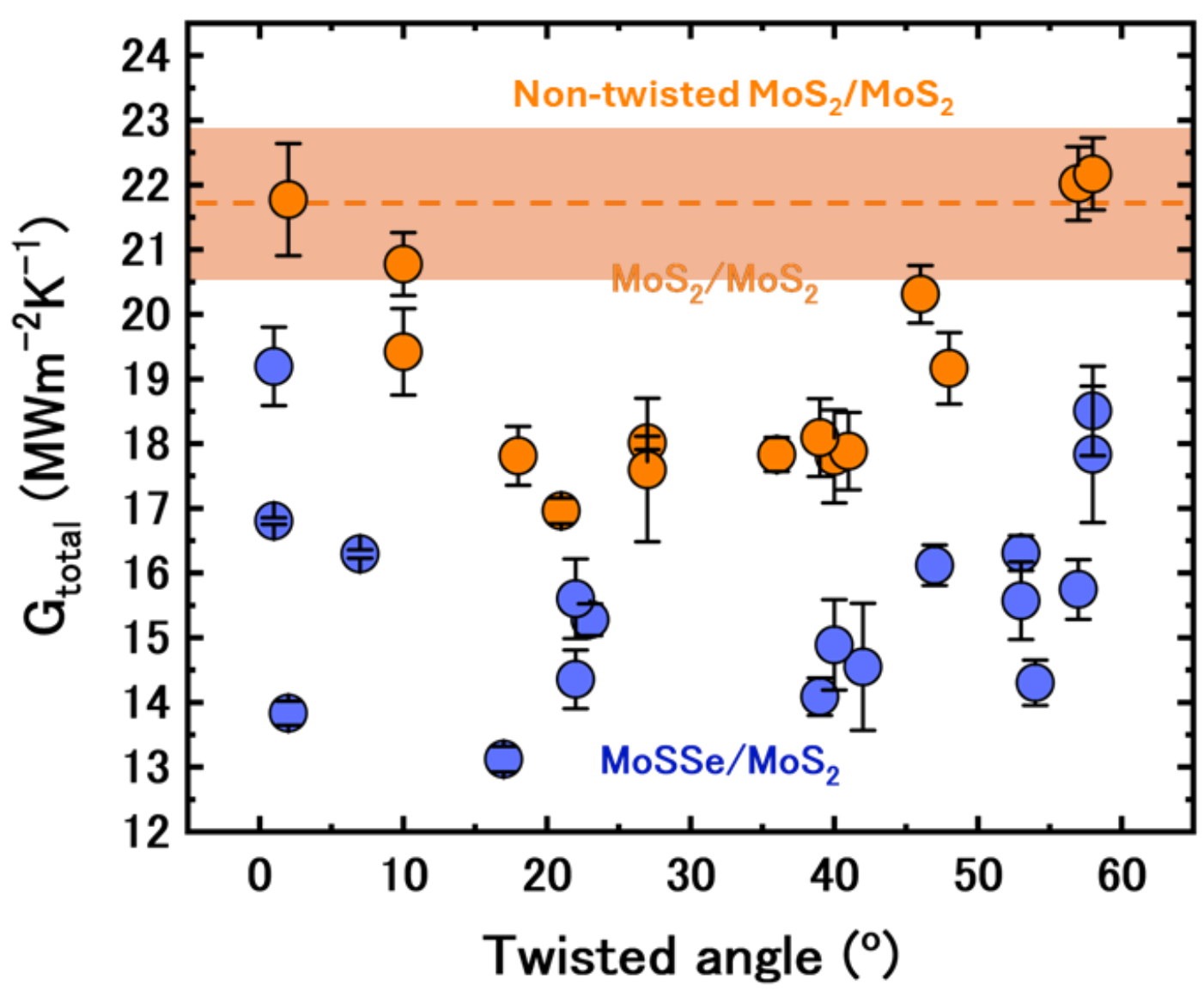

**Figure S2.** $G_{total}$ of bilayer $MoS_2$ and $MoSSe/MoS_2$.

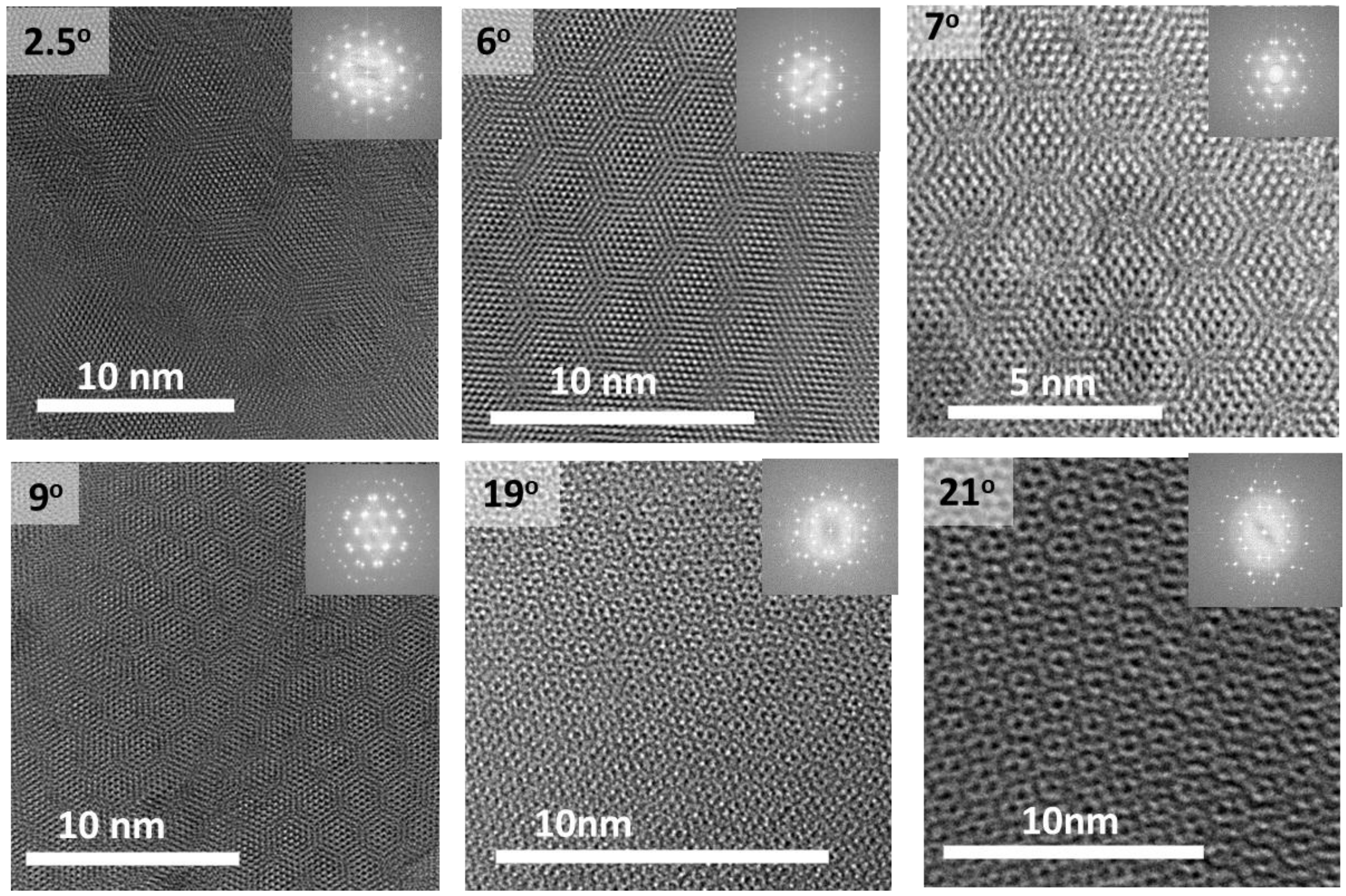

**Figure S3**. TEM image of bilayer $MoS_2$

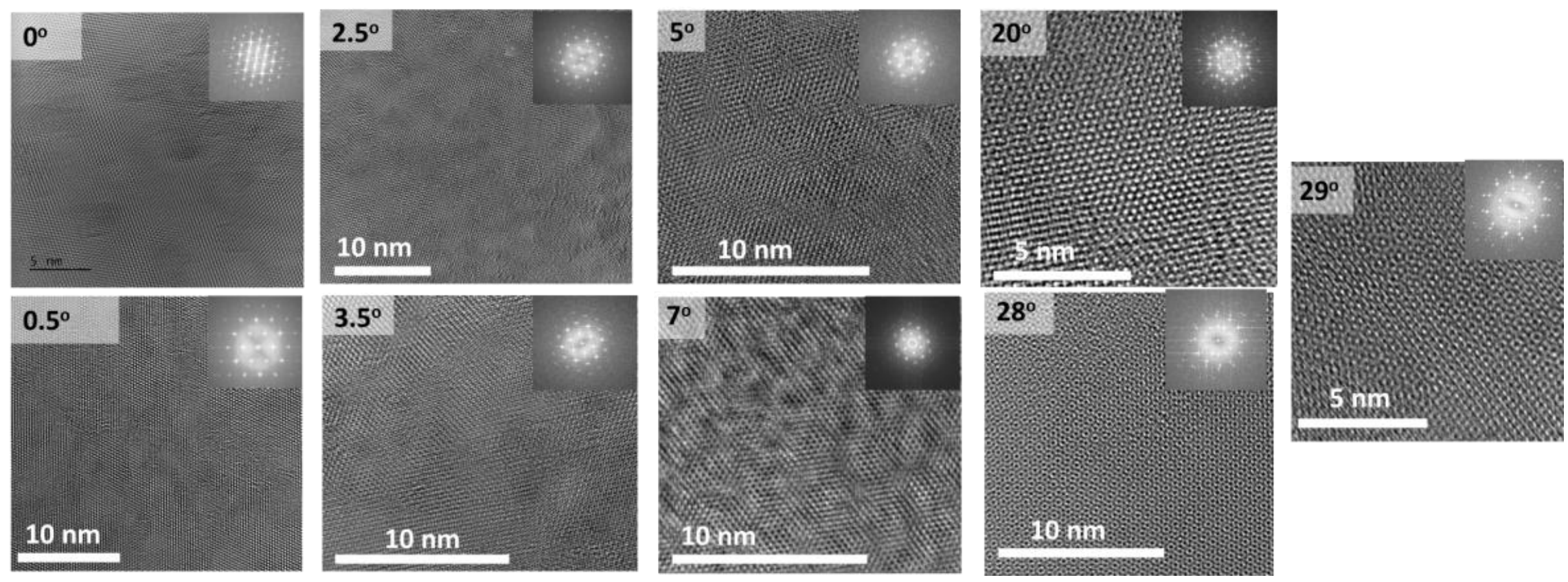

**Figure S4.** TEM image of bilayer $MoSSe/MoS_2$ with different twist angles.

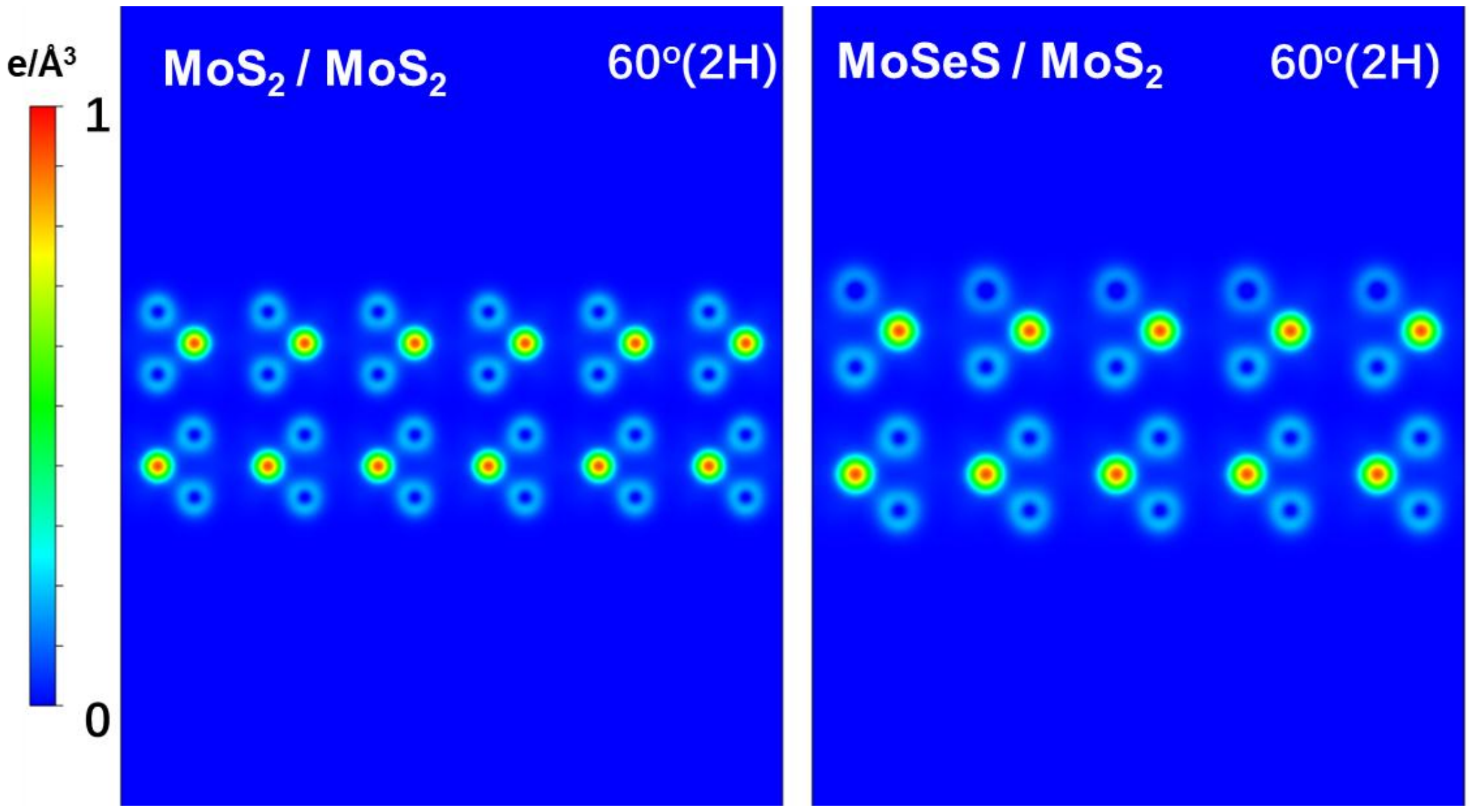

**Figure S5.** Charge density of $MoS_2$ and $MoSeS/MoS_2$ under a twist angle of $60^o$.

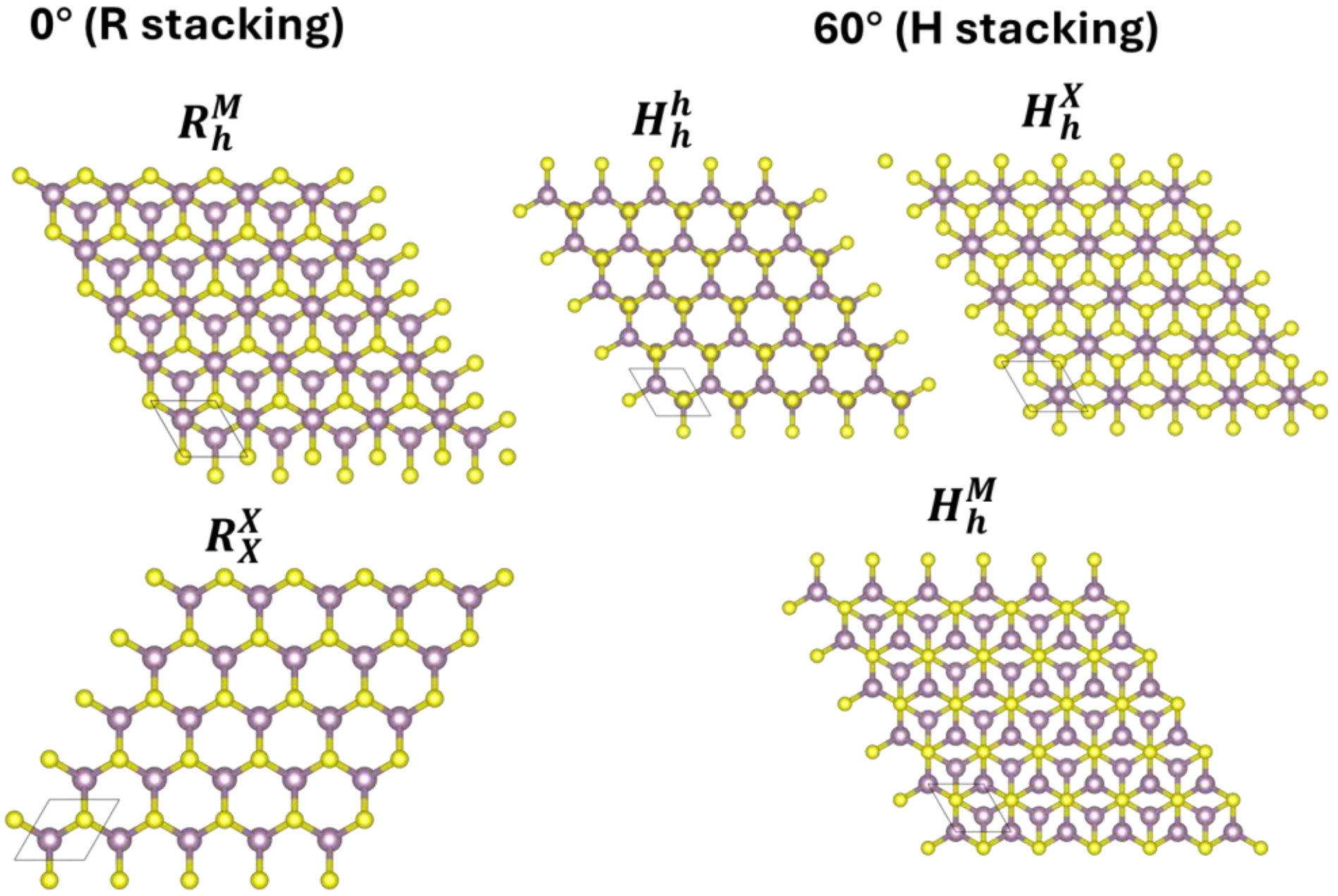

**Figure S6.** Structure of non-twisted $MoS_2/MoS_2$ with different stacking configuration; Top view.

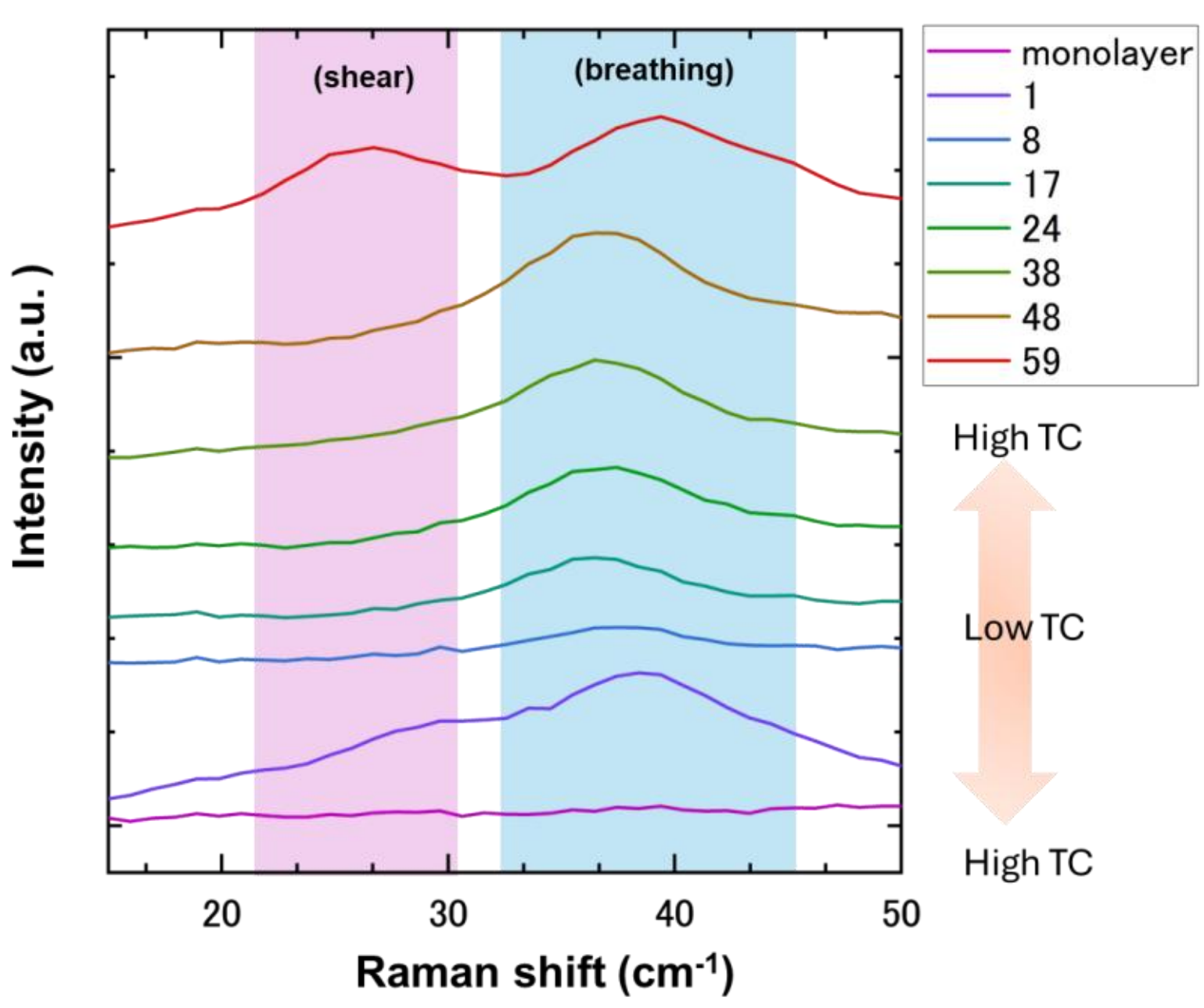

**Figure S7.** Low frequency Raman spectrum of $MoS_2$ and twisted bilayer $MoS_2$.

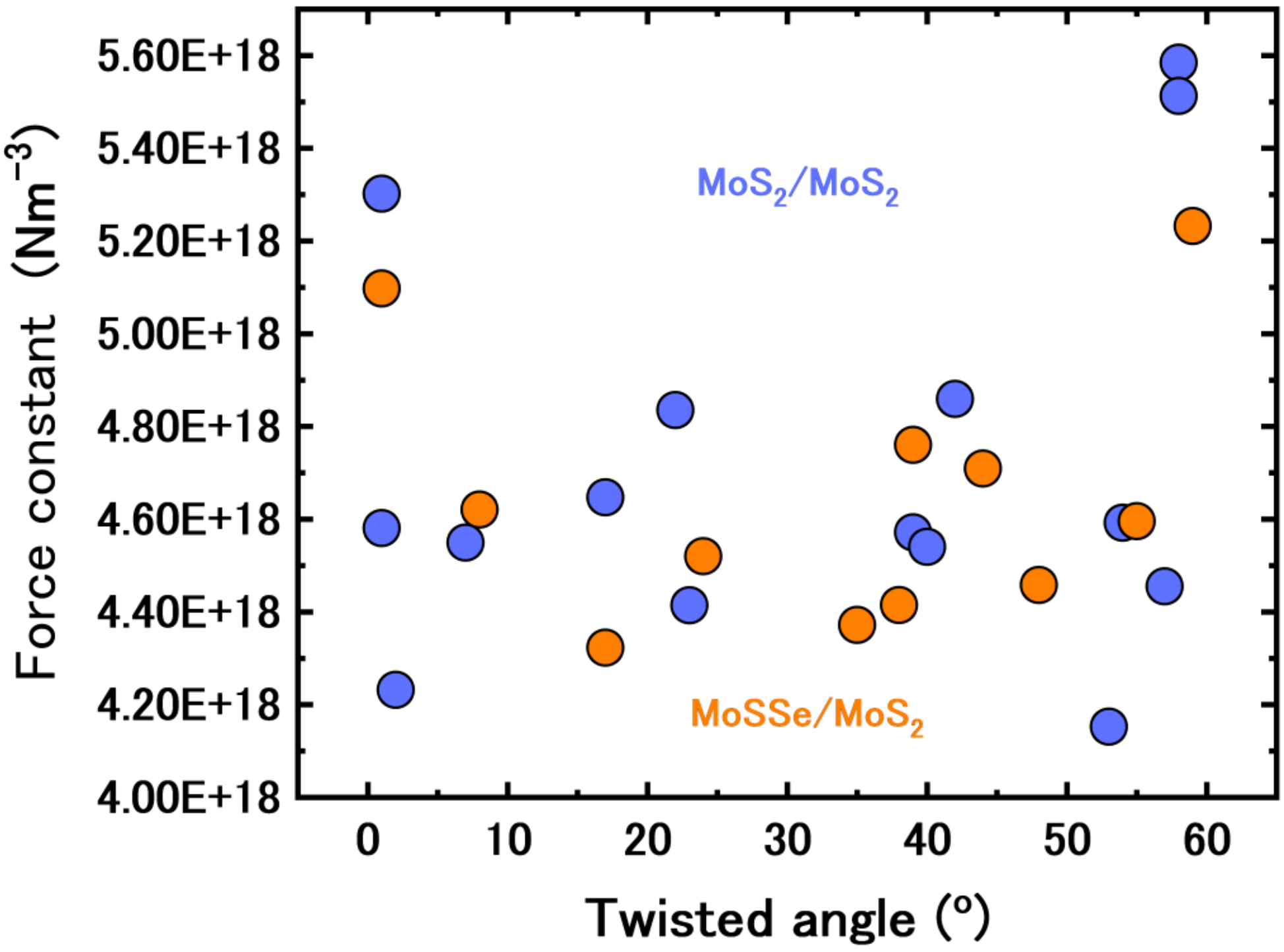

**Figure S8.** Out-of-plane force constant of twisted bilayer $MoS_2$ and $MoSSe/MoS_2$.

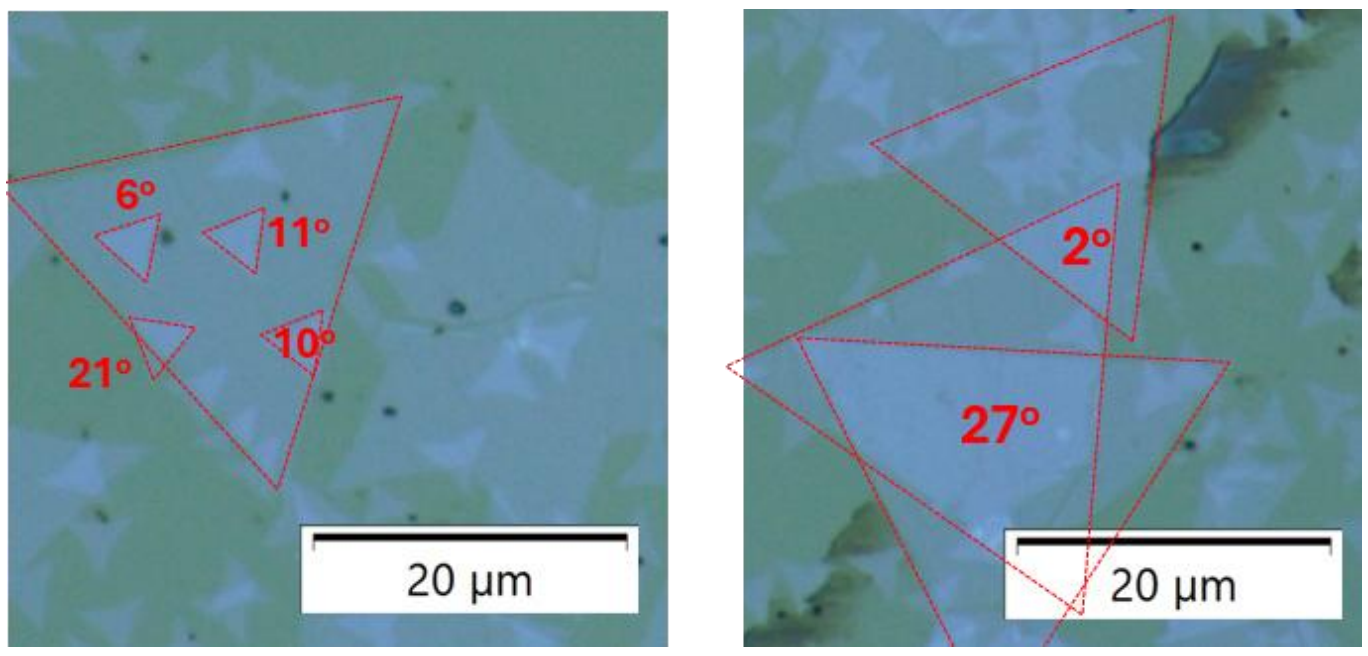

**Figure S9.** Twist-angle determination in twisted-bilayer TMD ($MoSSe/MoS_2$).

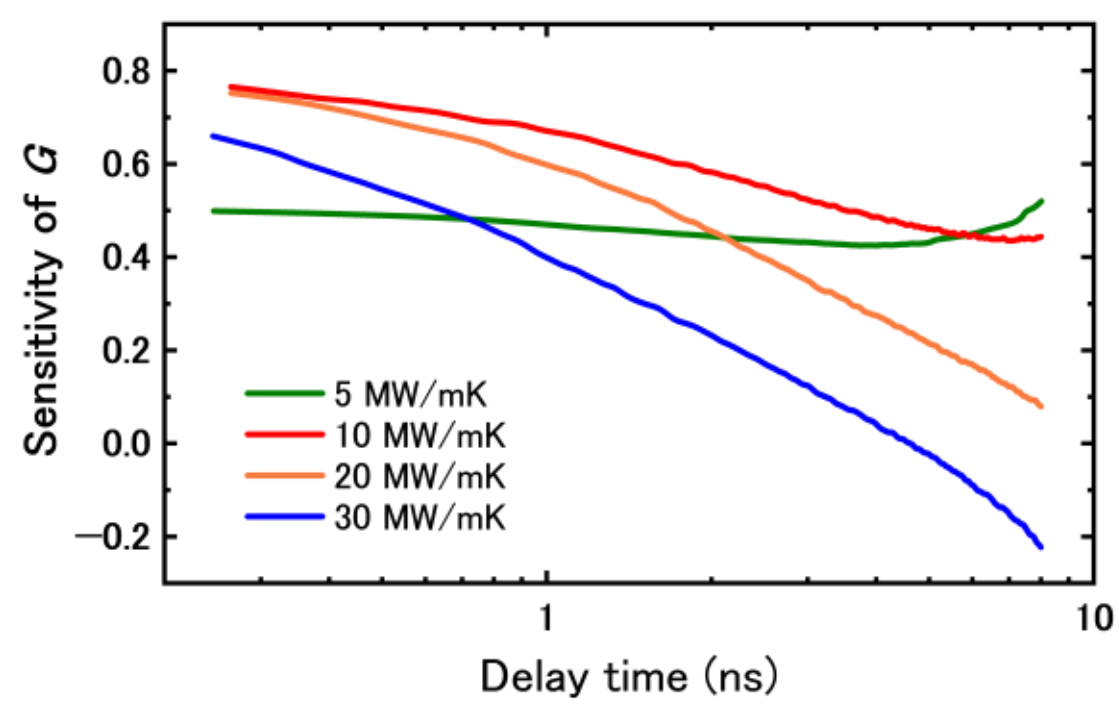

**Figure S10.** TDTR sensitivity calculation under TC of 5, 10, 20, 30 $MW/m^2K$, respectively.

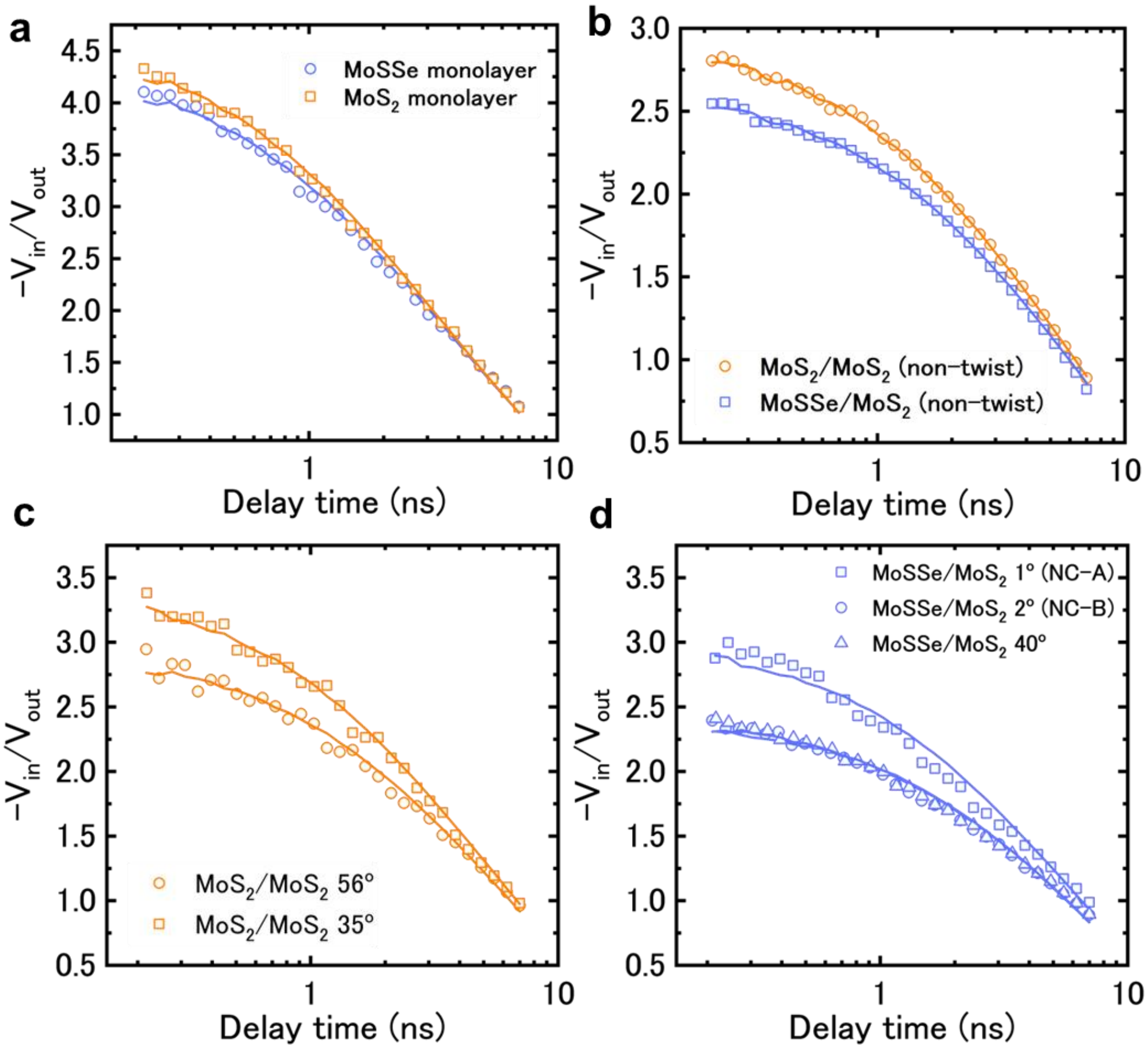

**Figure S11.** TDTR measured temperature decay (-Vin/Vout) signal of **a.** monolayer MoSSe and $MoS_2$; **b.** non-twist bilayer $MoS_2$ and MoSSe/$MoS_2$ synthesized from as-grown bilayer $MoS_2$; **c.** twisted bilayer $MoS_2$ with twist angle of 56° and 35°; **d.** twisted bilayer MoSSe/$MoS_2$ with twist angle of 1° (NC-A type), 2° (NC-B type) and 40°.

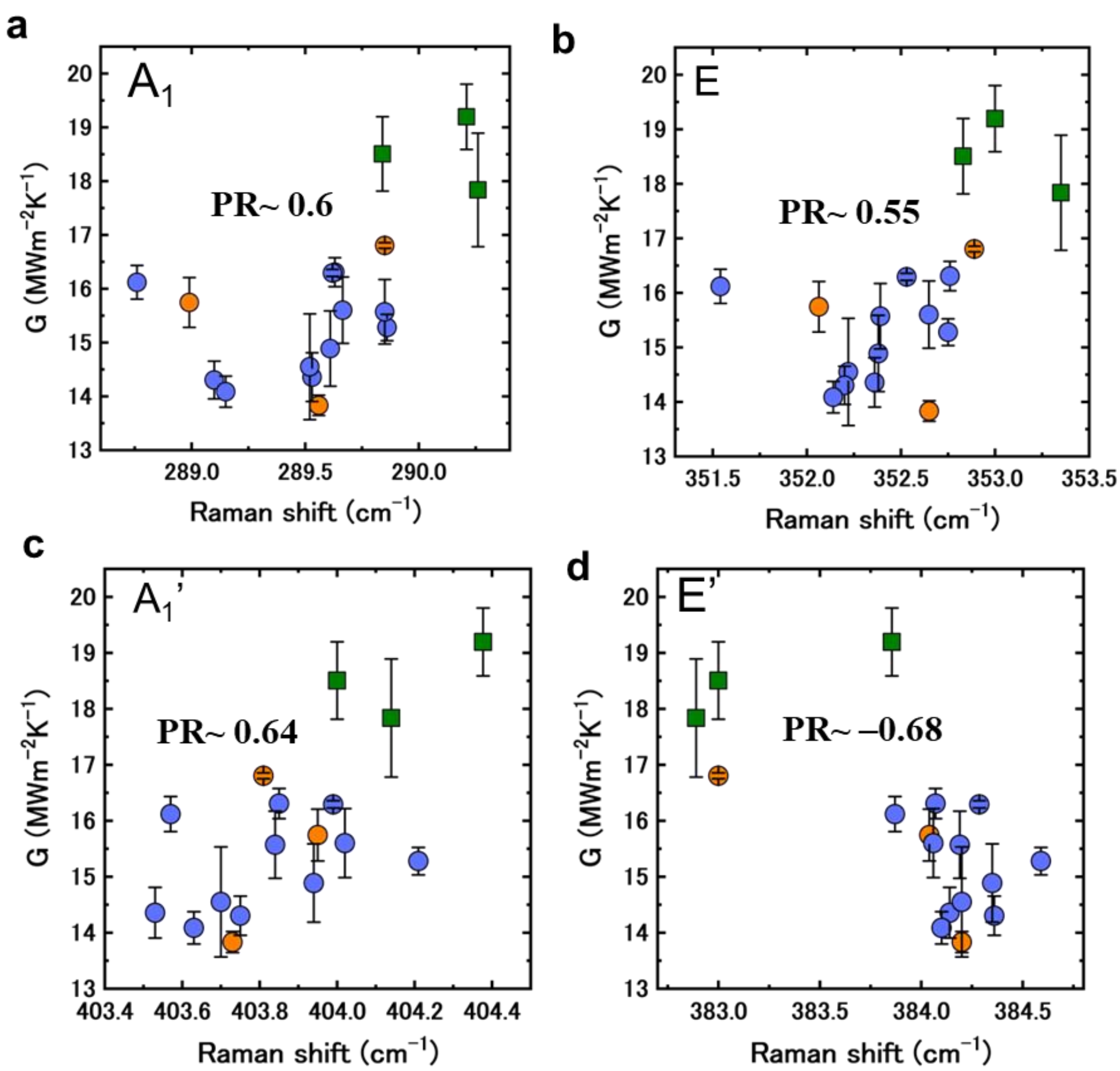

**Figure S12.** Correlation between *G* and high frequency Raman modes **a.** A1, **b.** E, **c.** $A_1$', **d.** E' of $MoSSe/MoS_2$ samples.

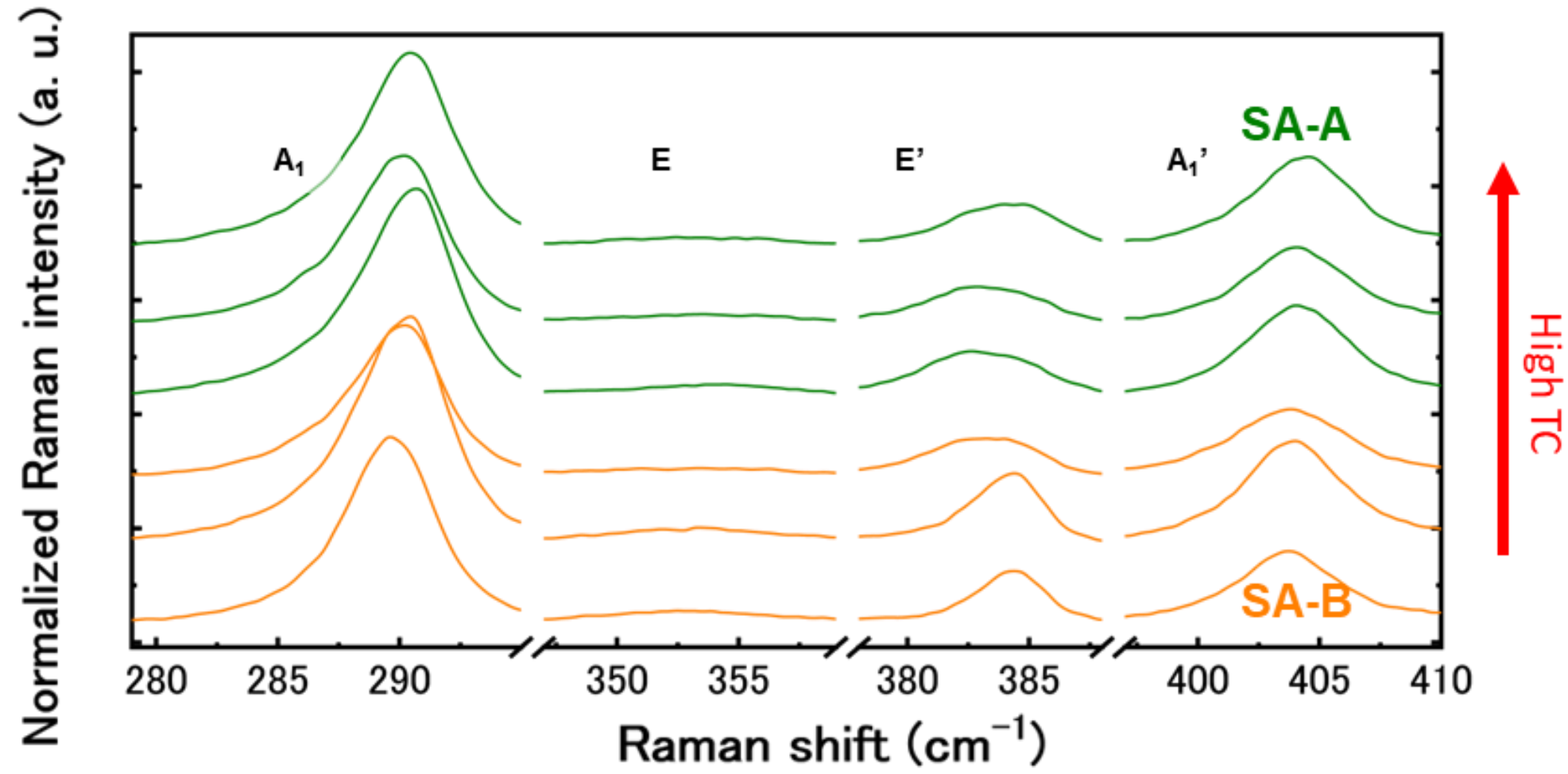

**Figure S13.** High frequency Raman spectra in NC-A and NC-B types of $MoSSe/MoS_2$ that within small twist angles.

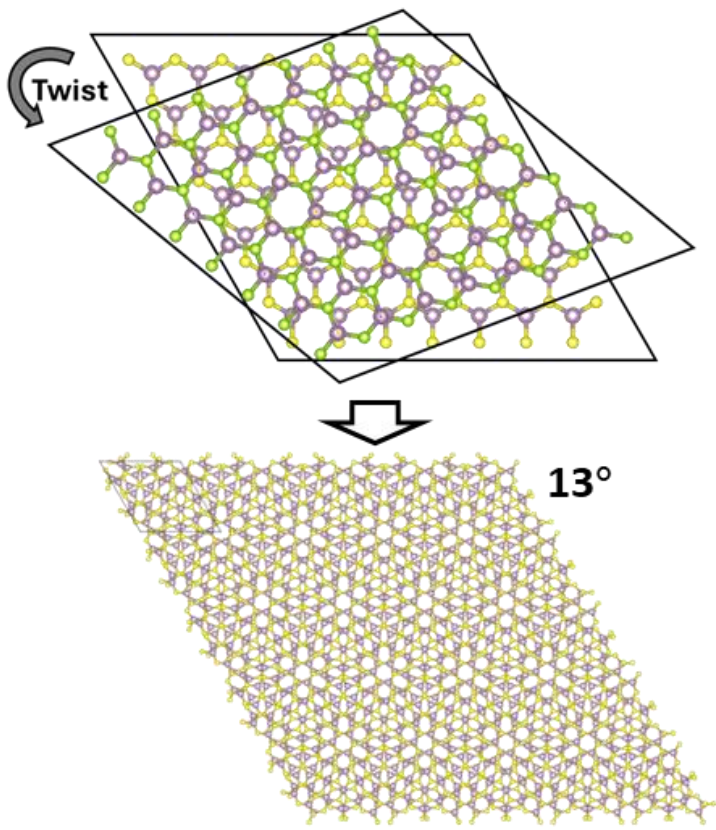

**Figure S14.** Schematic of building a twisted bilayer TMD structure.

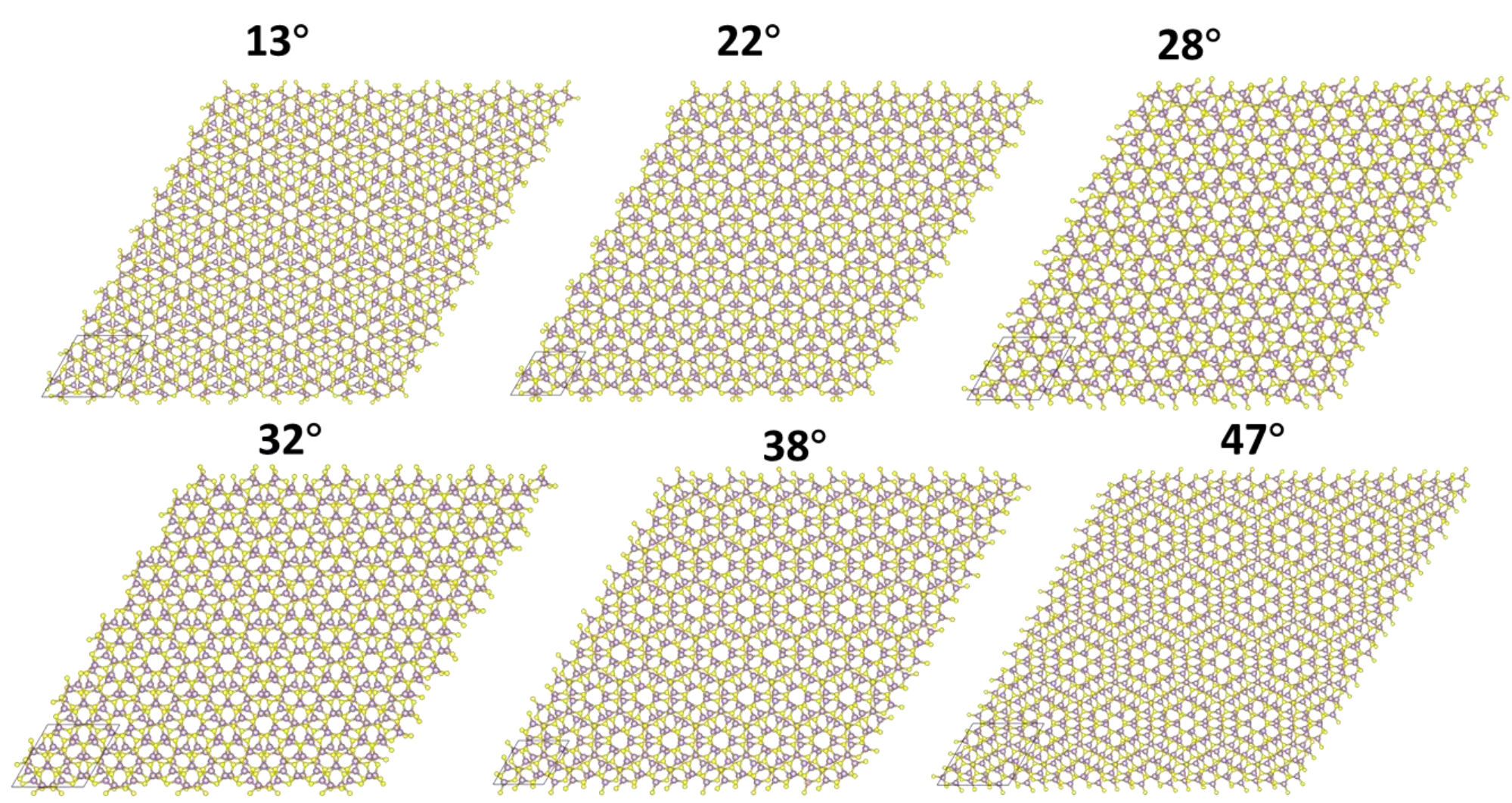

**Figure S15.** Structure of twisted $MoS_2/MoS_2$; Top view.

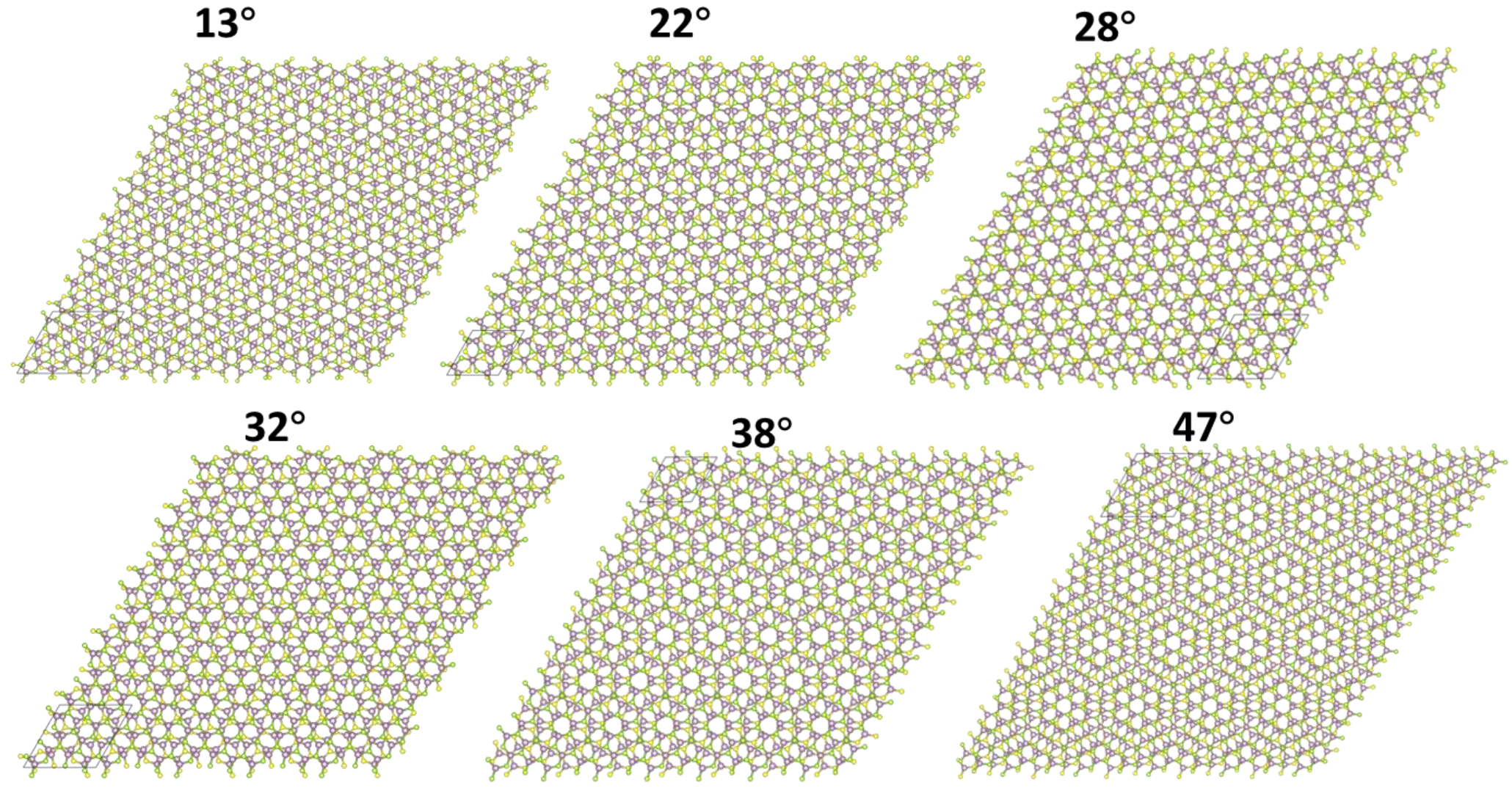

**Figure S16.** Structure of twisted $MoSSe/MoS_2$; Top view.

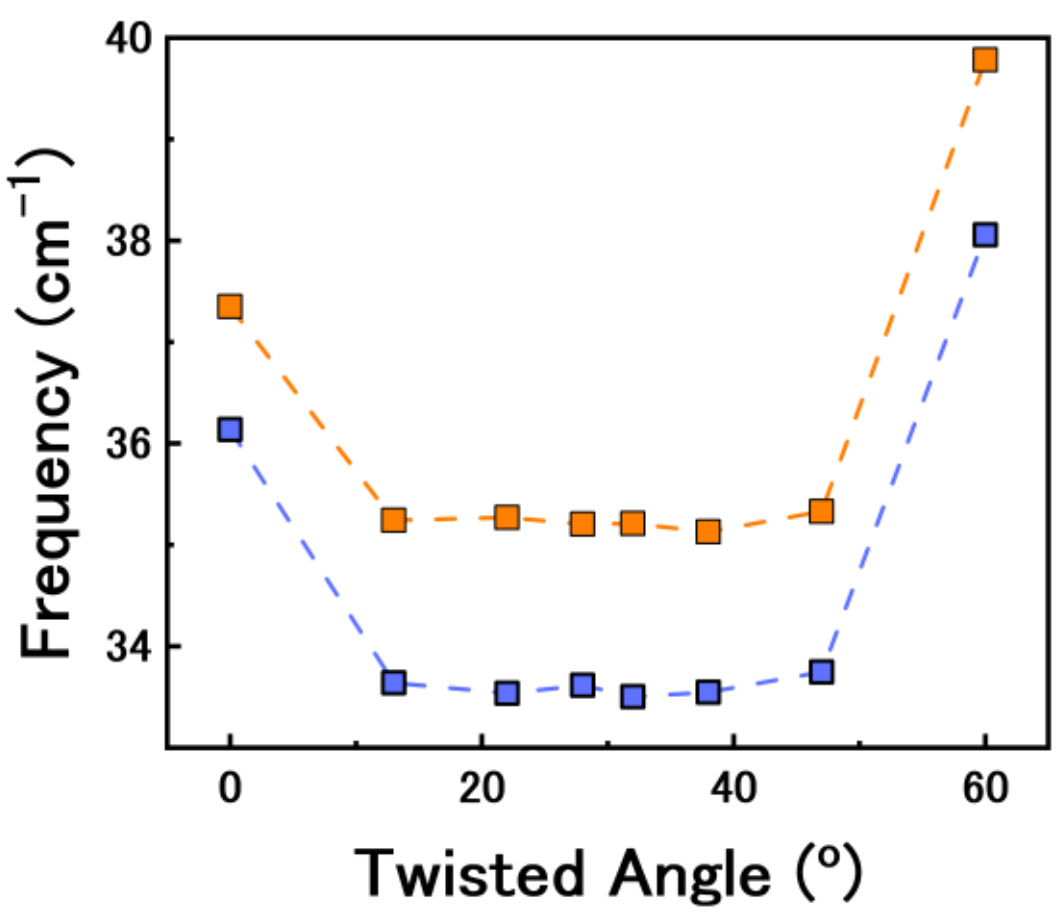

**Figure S17.** Breathing mode frequency of bilayer $MoS_2$ (purple) and $MoSSe/MoS_2$ (orange) under different twist angles calculated from DFT model.

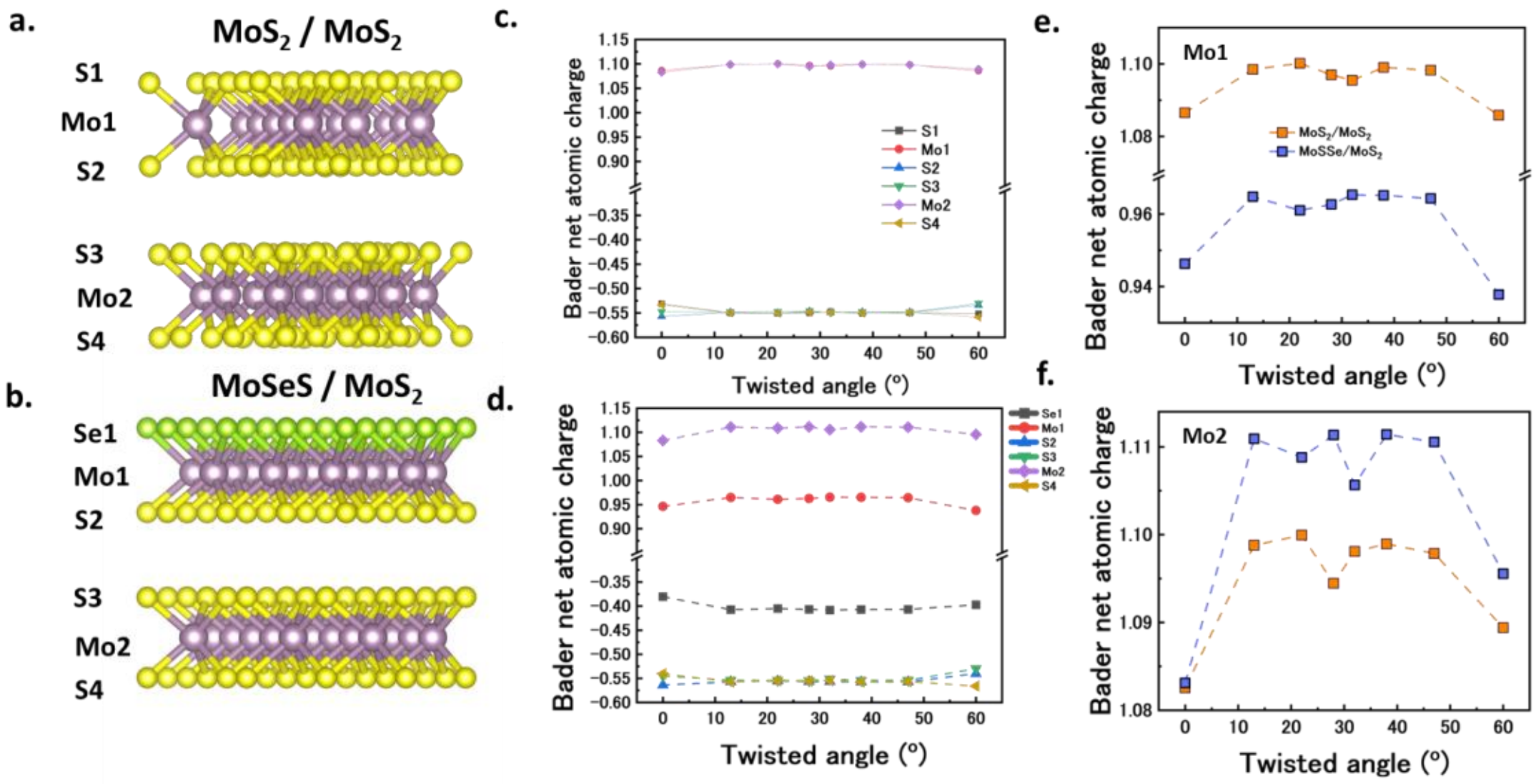

**Figure S18.** Schematic of **a.** bilayer $MoS_2$ and **b.** $MoSSe/MoS_2$; individual layers are numbered. Bader net atomic charge variation of **c.** bilayer $MoS_2$ and **d.** $MoSSe/MoS_2$ under different twist angles; the Bader net atomic charge of **e.** Mo1 and **f.** Mo2 in bilayer $MoS_2$ and $MoSSe/MoS_2$ under different twist angles.

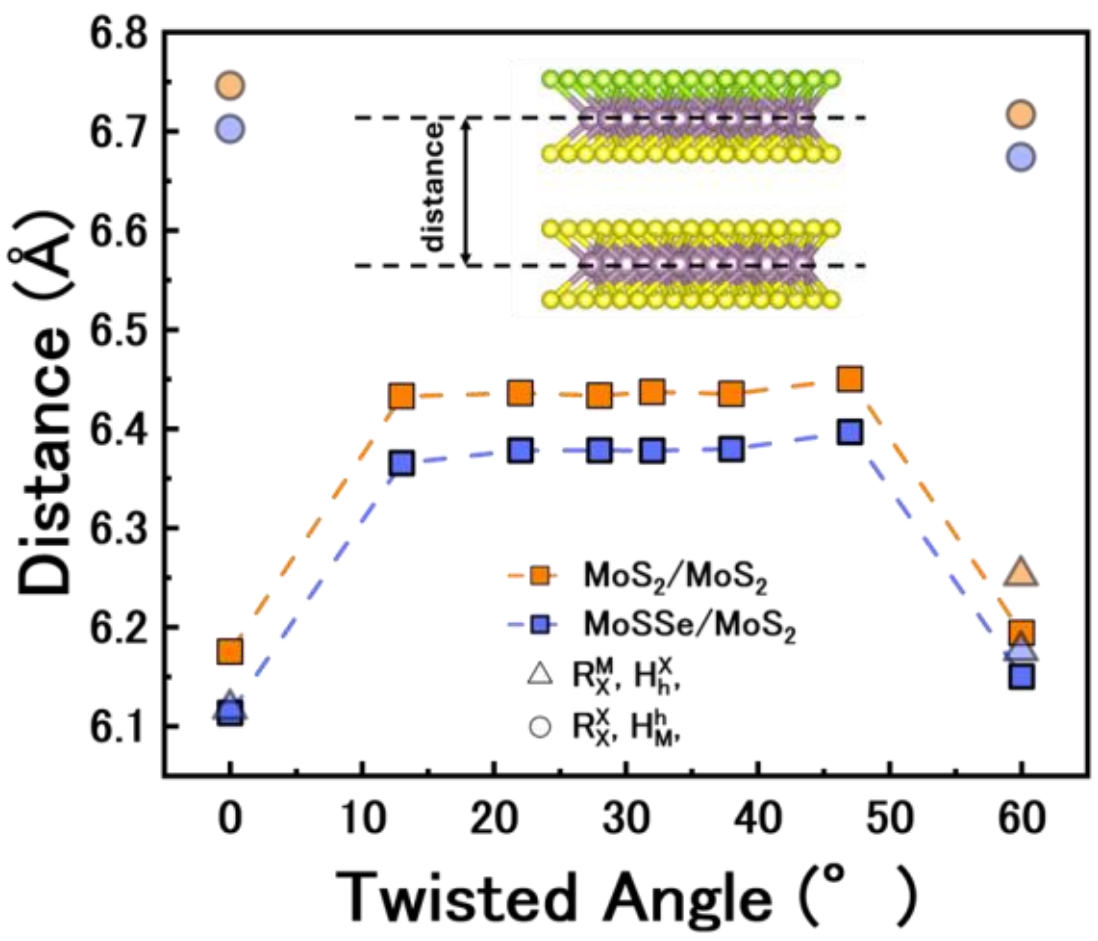

**Figure S19.** Interlayer distance of $MoS_2$ and $MoSeS/MoS_2$ under different twist angle of 60°.